\documentclass[twocolumn,twoside,journal]{IEEEtran} 

\usepackage{color}
\usepackage{cite}
\usepackage{graphicx}
\usepackage{dblfloatfix}
\usepackage{amsmath} 
\usepackage{amssymb}  
\usepackage{amsfonts}
\usepackage{mathrsfs}
\usepackage{mathtools}
\usepackage{lipsum}
\usepackage[thmmarks, amsmath]{ntheorem}
\usepackage[bbgreekl]{mathbbol}
\usepackage{multirow}
\usepackage{makecell}
\usepackage{booktabs}
\usepackage{enumitem}

\DeclareSymbolFontAlphabet{\mathbb}{AMSb}
\DeclareSymbolFontAlphabet{\mathbbl}{bbold}

\DeclareMathOperator{\diag}{diag}

\DeclareMathOperator{\blkdiag}{blkdiag}

\DeclareMathOperator{\diff}{d}

\DeclareMathOperator{\Null}{null}

\newcommand{\R}{{\mathbb R}}
\newcommand{\N}{{\mathbb N}}
\newcommand{\mc}{\mathcal}
\newcommand{\ddt}{\tfrac{\diff}{\diff \!t}}

\newcommand{\pv}{{\text{pv}}}
\newcommand{\dc}{{\text{dc}}}
\newcommand{\ac}{{\text{ac}}}
\newcommand{\g}{{\text{g}}}

\newcommand{\oo}{{\text{o}}}

\newcommand{\cc}{{\text{c}}}

\newcommand{\w}{{\text{w}}}

\newcommand{\gc}{{\mathfrak{g}}}
\newcommand{\fr}{{\text{r}}}

\newcommand{\mpp}{{\text{mpp}}}
	\newcommand{\sst}{{\text{ss}}}
\newcommand{\T}{{\mathsf{T}}}
	\newcommand{\pb}{{\text{bp}}}
	\newcommand{\zs}{{\text{zs}}}

\newtheorem{theorem}{Theorem}
\newtheorem{lemma}{Lemma}
\newtheorem{proposition}{Proposition}

\newtheorem{condition}{Condition}
\newtheorem{example}{Example}
\newtheorem{corollary}{Corollary}

\newcommand{\ad}{{{\text{c}_\text{ac}}}}
\newcommand{\da}{{{\text{c}_\text{dc}}}}

\usepackage{xcolor}

\title{\LARGE \bf  Universal dual-port grid-forming control: bridging the gap between grid-forming and grid-following control}

\author{Irina Suboti\'c and Dominic Gro\ss{} \thanks{I. Suboti\'c is with the Automatic Control Laboratory at ETH Z\"urich, Switzerland, D. Gro\ss{} is with the Department of Electrical and Computer Engineering at the University of Wisconsin-Madison, USA; e-mail:subotici@ethz.ch,  dominic.gross@wisc.edu}}

\begin{document}
	\maketitle
	\begin{abstract} 
		We analyze a dual-port grid-forming (GFM) control for power systems containing ac and dc transmission, converter-interfaced generation and energy storage, and legacy generation. To operate such a system and provide standard services, state-of-the-art control architectures i) require assigning grid-following (GFL) and GFM controls to different converters, and ii) result in highly complex system dynamics. In contrast, dual-port GFM control (i) subsumes common functions of GFM and GFL controls in a simple controller, ii) can be applied to a wide range of emerging technologies independently of the network configuration, and iii) significantly reduces system complexity. In this work, we provide i) an end-to-end modeling framework that allows to model complex topologies through composition of reduced-order device models, ii) an in-depth discussion of universal dual-port GFM control for emerging power systems, and iii) end-to-end stability conditions that cover a wide range of network topologies, emerging technologies, and legacy technologies. Finally, we validate our findings in detailed case studies.
	\end{abstract}
\section{Introduction}
Electrical power systems are transitioning from fuel-based legacy synchronous generators (SGs), whose physical properties (e.g., inertia) and controls (e.g., governor) form the foundation of today's system operation and analysis, to converter-interfaced resources such as renewable generation, energy storage systems, and high voltage direct current (HVDC) transmission. This large-scale integration of converter-interfaced resources results in significantly different system dynamics that jeopardize stability and reliability~\cite{MDH+18}.
	
Today, the majority of converter-interfaced resources uses GFL control to, e.g., maximize the energy yield of renewables or minimize HVDC transmission losses. This approach typically relies on a phase-locked loop (PLL) for synchronization and assumes that a stable ac voltage waveform (i.e., frequency and magnitude) at the point of interconnection is guaranteed by, e.g., the presence of SGs. While GFL power converters can provide typical ancillary services (e.g., primary frequency control), dynamic stability of power system can rapidly deteriorate as the share of GFL resources increases \cite{MDH+18,MBP+2019}. 
	
To resolve this issue, GFM converters that impose a stable ac voltage at their terminal and self-synchronize are envisioned to be the cornerstone of future power systems \cite{MBP+2019}. While prevalent GFM controls such as active power - frequency ($P_\ac\!-\!f$) droop control \cite{CDA93}, virtual synchronous machine control \cite{DSF2015}, and (dispatchable) virtual oscillator control \cite{SG+20}, provide fast and reliable grid support~\cite{MBP+2019}, they may destabilize the system if the resource interfaced by the converter reaches its power generation limits~\cite{TGA+20}.

While the classification into GFM and GFL control is commonly applied to the point of connection with the ac grid (i.e., ac-GFM and ac-GFL) it is also useful to characterize the dc terminal behavior of dc/ac voltage source converters (VSCs) for renewable integration and HVDC transmission, i.e., dc-GFL assumes a stable dc voltage while dc-GFM stabilizes the VSC dc terminal \cite{GSP+21,LSG22,SGP+23}. Most of the existing literature focuses on ac-GFM/dc-GFL and dc-GFM/ac-GFL control and treats these concepts as mutually exclusive and complementary, i.e., operating emerging power systems requires assigning ac-GFL/dc-GFM and ac-GFM/dc-GFL controls to different converters (e.g., to support maximum power point tracking (MPPT) \cite{LSG22}, high-voltage ac (HVAC) and reliably operate HVDC transmission \cite{SGP+23}). This approach results in complex, heterogeneous system dynamics that introduces significant challenges in system operation, e.g., assigning GFM/GFL roles is non-trivial and no control configuration may be stable for all relevant operating points \cite{SGP+23} and/or available reserves \cite{TGA+20}. Finally, to the best of our knowledge, no analytical stability conditions are available for systems containing ac-GFM/dc-GFL VSCs, ac-GFL/dc-GFM VSCs, and legacy devices.

Broadly speaking, most of the existing works on analytical stability analysis of ac power systems with ac-GFM VSCs neglect the dc terminal (see, e.g., \cite{SG21} and references therein for details). In contrast, dc voltage - frequency ($v_\dc\!-\!f$) droop inspired by machine emulation control~\cite{CBB+2015} is analyzed in \cite{MDPS+17,TGA+20} assuming proportional dc voltage control through its dc source. Moreover, the lead-lag control proposed in~\cite{HHH+18} can be understood as an approximation of the PD control \eqref{eq:control.law} analyzed in this work. While variants of $v_\dc\!-\!f$ droop control have been proposed before, to the best of the authors' knowledge, no  end-to-end stability results that account for the dynamics of (renewable) generation are available.

An often overlooked feature of $v_\dc\!-\!f$ droop control is its ability to bridge the gap between ac-GFM/dc-GFL and ac-GFL/dc-GFM control by simultaneously controlling ac frequency and dc voltage (ac-GFM/dc-GFM)~\cite{SG21,GSP+21,LSG22}. This allows unifying functions of ac-GFL/dc-GFM (e.g., MPPT) and ac-GFM/dc-GFL (e.g., primary frequency control) in a dual-port GFM control that reduces complexity by using the same control independently of the network configuration \cite{GSP+21} and VSC power source \cite{SG21}. In contrast to standard ac-GFM/dc-GFL and ac-GFL/dc-GFM control, dual-port GFM control provides bidirectional grid support functions whose direction across the dc/ac interface autonomously adapts to the system topology, operating point, and control reserves. For instance, for photovoltaics (PV) operating at their maximum power point (MPP), the dc voltage is stabilized through controlling the ac power injection. In contrast, curtailed PV stabilizes the dc voltage which in turn stabilizes the ac frequency. Combining $v_\dc-f$ and $P_\ac-f$ droop control results in so-called power-balancing dual-port GFM control for which analytical stability conditions for systems containing ac and dc transmission, renewables, SGs, and synchronous condensers (SCs) are available \cite{SG21}.

Despite these appealing features, power-balancing dual-port GFM control has several conceptual drawbacks. Specifically, significant oscillations may occur during transients because the transient and steady-state response cannot be tuned separately \cite{GSP+21}, post-disturbance steady-state frequencies are generally not synchronous for ac systems interconnected through HVDC \cite{SG21}, and  $P_\ac\!-\!f$ droop may result in a counter-intuitive response when post-contingency power flows significantly differ from the VSC power set-point \cite{GSP+21}. Moreover, for some system topologies, stability conditions still hinge on the presence of controllable power sources on VSC dc terminals \cite{SG21}.

Instead, the main contribution of this is a rigorous end-to-end stability analysis of a universal dual-port GFM control based on proportional-derivative (PD) $v_\dc\!-\!f$ droop control \cite{GSP+21,LSG22} that retains all features of power-balancing dual-port GFM control while overcoming its conceptual limitations. In particular, we show that the transient and steady-state response can be tuned separately and that, for typical HVDC systems, ac systems interconnected through HVDC admit a quasi-synchronous steady-state. Moreover, the disturbance response of universal dual-port GFM control is well-behaved even if post-contingency power flows significantly differ from the dispatch (i.e., generator power and VSC dc voltage).

We propose to use dual-port GFM control for i) common renewable sources and plant architectures to enable both approximate maximum power point tracking as well as standard grid-support function, and ii) to enable transparent grid-support functions through HVDC. Simplified stability conditions  are available for a single VSCs \cite{GSP+21} and a single back-to-back connected wind turbine \cite{LSG22}. Our main technical contribution is a comprehensive end-to-end stability and steady-state analysis of universal dual-port GFM control that establishes i) stability and ii) steady-state analysis for a wide range of emerging systems including legacy synchronous generation, converter interfaced renewable generation and HVDC links. These results provide deep engineering insights. Broadly speaking, we show that there have to be sufficiently many devices with frequency control reserves and that, e.g., synchronous condensers (SCs) cannot be arbitrarily connected to ac grids. Moreover, our theoretical results clarify the relationship between system stability, curtailment of renewables, and the overall system topology. Notably, compared to power-balancing dual-port GFM control \cite{SG21}, universal dual-port GFM control does not require additional signals (e.g.,  active power and active power set points) beyond the converter dc-link voltage. Crucially, omitting power setpoints allows for i) autonomous rebalancing of post-disturbance power flows, and ii) separate tuning of the steady-state and transient response. To make the control more practical, we propose a derivative-free implementation of universal dual-port GFM control. We illustrate that universal dual-port GFM control provides autonomous bidirectional grid support (i.e., primary frequency control) through asynchronous interconnections (e.g., HVDC). Finally, to illustrate the analytical results two case studies are used. A case study based on the IEEE 9-bus system and detailed switched converter models is used to illustrate the response of a solar PV system and a PMSG WT. Finally, averaged converter models are used to illustrate the results in a large-scale system containing renewable generation, HVDC, and legacy generation.
\subsubsection*{Notation}
Given a matrix $A$, $A\succcurlyeq0$ ($A\succ0$) denotes that $A$ is symmetric and positive semidefinite (definite).  The identity matrix of dimension $n$ is denoted by $I_n$. Matrices of zeros of dimension $n \times m$ are denoted by $\mathbbl{0}_{n\times m}$. Column vectors of zeros and ones of length $n$ are denoted by $\mathbbl{0}_{n}$ and $\mathbbl{1}_{n}$. Given $x\in\R^n$ and $y\in\R^m$, we define $(x,y) \coloneqq [x^\mathsf{T}\; y^\mathsf{T}]^\mathsf{T} \in \R^{n+m}$. The cardinality of a set $\mc X \subset \mathbb{N}$ is denoted by $|\mc X|$. Deviations of a signal from its operating point $(\cdot)^\star$ is denoted via $(\cdot)_\delta$. Frequently used variables are defined in Tab.~\ref{table:nomenclature} while the remaining variables are defined when needed. 
\begin{table}	
	\centering
	\caption{Nomenclature}\label{table:nomenclature}
			\setlength\tabcolsep{0.1pt}
			\bgroup
			\def\arraystretch{1.7}%
	\begin{tabular}{c p{63mm}}
		\toprule

		$\mc N_\ac$, $\mc N_\cc$, $\mc N_\dc$ & Node sets collecting ac nodes, dc/ac nodes, and dc nodes\\
	\hline
	$\mc E_\ac$, $\mc E_\dc$
		& Edges corresponding to ac and dc connections \\
		\hline
		$\mc G_N$, $\mc N_N$, $\mc E_N $ & System graph, nodes, and edges \\
		\hline
		$\mc G_\ac$, $\mc G_\dc$ & AC and dc graphs \\
		\hline
		$N_\ac,N_\dc \in \N$  & Number of maximal connected ac and dc subgraphs \\
		\hline
		$\mc N_\ac^i$, $\mc N_\ad^i$ & AC and ac/dc nodes of the $i$th 
		 connected ac subgraph\\
		\hline
		$\mc N_\dc^i$, $\mc N_\dc^i$ & DC and dc/ac nodes of the $i$th connected dc subgraph \\
		\hline
		$\mc E_\ac^i$, $\mc E_\dc^i$ &  Edges of the $i$th maximal connected ac and dc subgraph\\
		\hline
		$\mc G_\ac^i$, $\mc G_\dc^i$ & $i$th maximal connected ac and dc subgraph \\
		\hline
		$(\cdot)^\star$, $(\cdot)_{\delta,l}$ &  Setpoint and deviation of a variable from its setpoint \\
		\hline
		{$\omega_l$, $\omega_l^\star$, $\omega_{\delta,l}$} & AC voltage frequency, setpoint, and deviation\\
		\hline
		{$\theta_l$, $\theta_l^\star$, $\theta_{\delta,l}$} & AC voltage phase angle, setpoint, and deviation  \\
		\hline
		{$\eta_m$, $\eta_{m}^\star$, $\eta_{\delta,m}$} & AC phase angle difference, setpoint, and deviation \\
		\hline
		{$V_l$, $V_l^\star$, $V_{\delta,l}$} & AC voltage magnitude, setpoint, and deviation \\
		\hline
		{$v_l$, $v_l^\star$, $v_{\delta,l}$} & DC voltage, setpoint, and deviation \\
		\hline
		{$P_l$, $P_l^\star$, $P_{\delta,l}$} & Power generation, setpoint, and deviation \\
		\hline
		$P_{\ac,l}$, $P_{\ac,l}^\star$, $P_{\delta,\ac,l}$ & AC power injection, setpoint, and deviation \\
		\hline
		$P_{\dc,l}$, $P_{\dc,l}^\star$, $P_{\delta,\dc,l}$  & DC power injection, setpoint, and deviation \\
		\hline
		$P_{d_\ac,l}$, $P_{d_\dc,l}$ & AC and dc load perturbation \\
		\hline
		$k_{\g,l}$ & Sensitivity of controllable ac and dc generation\\
		\hline
		$k_{\pv,l}$ & Sensitivity of PV \\
		\hline
		$k_{\w,l}$, $k_{\beta,l}$ & WT sensitivity to rotor speed and blade pitch angle \\
		\hline
		$P_{\fr,\delta}$, $P_{\zs,\delta}$ & Vector of $P_{\delta,l}$ with $k_{\g,l}>0$ and $k_{\g,l}=0$\\
		\hline
		$P_l^\mpp$ & MPP of renewable generation \\
		\hline 
		$v_l^\mpp$ & MPP dc voltage of PV\\
		\hline
		$\omega_l^\mpp$ & MPP rotor speed of WT \\
		\hline 
		$b_{kl}^\ac$, $g_{kl}^\dc$ & AC line suceptance and dc line conductance\\
		\hline 
		$\mc N_\fr$, $\mc N_\zs$ & Nodes connected to generation with $k_{\g,l}>0$ and  $k_{\g,l}=0$ \\
		\hline
		 $\mc N_\pv$ & Nodes connected to PV with $k_{\pv,l}>0$ \\
		\hline 
		$\mc N_\zs$ & Nodes connected to WT with $k_{\w,l}>0$ \\
		\hline
		$L_\dc$ & Laplacian matrix of the dc graph \\
		\hline
		$ B_\ac$ & Incidence matrix of the ac graph\\
		\hline
		$k_{p,l}$, $k_{\omega,l}$ & control gains of dual-port GFM control \\
		\hline
		$\mc N_{\ac^\fr}^i$ & AC nodes with $k_{\g,l}>0$ or $k_{\w,l}>0$ in the $i$th connected ac subgraph \\
		\hline 
		$\mc N_{\ac^\oo}^i = \mc N_\ac^i\setminus \mc N_{\ac^\fr}^i$ & remaining ac nodes in the $i$th subgraph \\
		\hline  
		$\mc N_{\ad^\fr}^i$ & AC/DC nodes connected to a dc network with dc voltage control\\
		\hline
		$\mc N_{\ad^\oo}^i = \mc N_\ac^i\setminus \mc N_{\ad^\fr}^i$ & remaining ac/dc nodes in the $i$th subgraph \\
		\bottomrule  
	\end{tabular}
	\egroup	
\end{table}

\section{Power system modeling}	\label{sec:modeling}
This section briefly summarizes linearized device and network models and illustrates how complex power systems encompassing wide range of different technologies (e.g., solar PV, wind generation, HVDC links, etc.) can be modeled by combining individual device and network models through a graph-based modeling framework. 
\subsection{Illustrative example} \label{sec:illustrative.example.ac.dc}
\begin{figure}[t]
	\centering
	\includegraphics[width=0.5\textwidth]{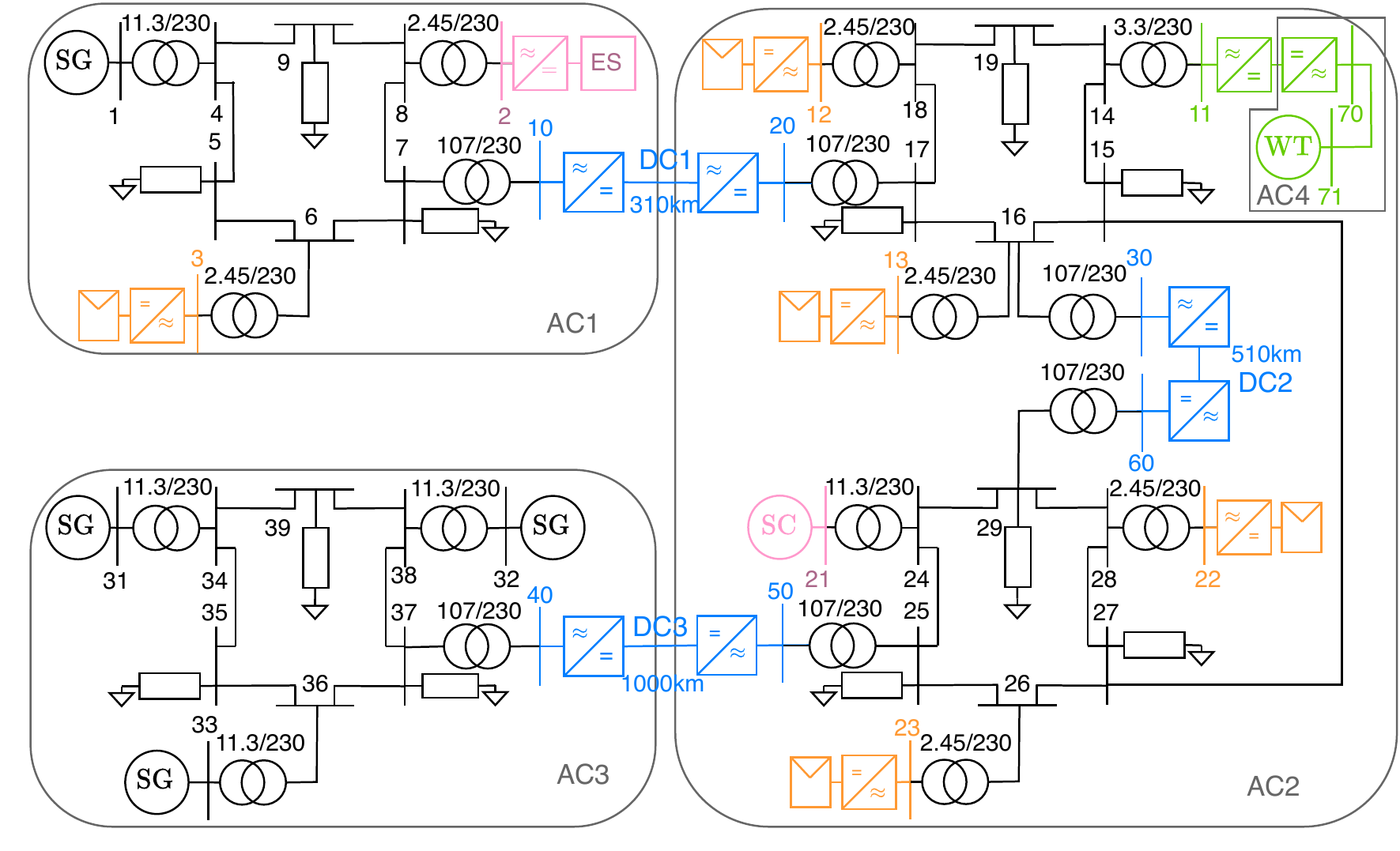} 
	\caption{Hybrid ac/dc power grid with renewable generation, legacy generation, HVDC links, battery storage, and a synchronous condenser.\label{fig:grid}}
\end{figure}
The system depicted in Fig.~\ref{fig:grid}, is used throughout the manuscript to illustrate our modeling framework, stability conditions and case study. The system consists of three bulk systems, each based on IEEE~9-bus systems. The system AC~1 contains a single-stage PV plant (orange), a battery energy storage system (pink), and legacy SGs. AC~2 consists of two systems interconnected through ac and dc transmission and includes four single-stage PV plants (orange), a synchronous condenser (pink), and a permanent magnet synchronous generation (PMSG) wind turbine (WT) (green). The system AC~3 only contains legacy SGs. Finally, three HVDC links (blue) connect the AC systems AC~1, AC~2, and AC~3. We emphasize that the ac and dc connections internal to the PMSG WT and PV plants are also modeled as ac and dc networks. 

Next, we illustrate the level of complexity that prevalent ac-GFM/dc-GFL and ac-GFL/dc-GFM controls introduce. Namely, operation of the system in Fig.~\ref{fig:grid} may require up to six variants of conventional VSC ac-GFM/dc-GFL and ac-GFL/dc-GFM control strategies and, depending on the desired operating mode and functions (e.g., MPPT, grid-support) of renewables, online switching between controls may be required. Moreover, HVDC links require a carefully designed combination of  ac-GFM/dc-GFL and ac-GFL/dc-GFM control \cite{SGP+23}. Instead, we proposed to reduce complexity, enable interoperability between various devices, and allow for end-to-end analysis, by using the same universal control (see Sec.~\ref{sec:control}) for all VSCs irrespective of the power generation technology or network topology.
\subsection{Network model} \label{sec:power.network.topology}
	\begin{figure}[t]
	\centering
	\includegraphics[trim=0 5mm 0 0, clip, width=0.5\textwidth]{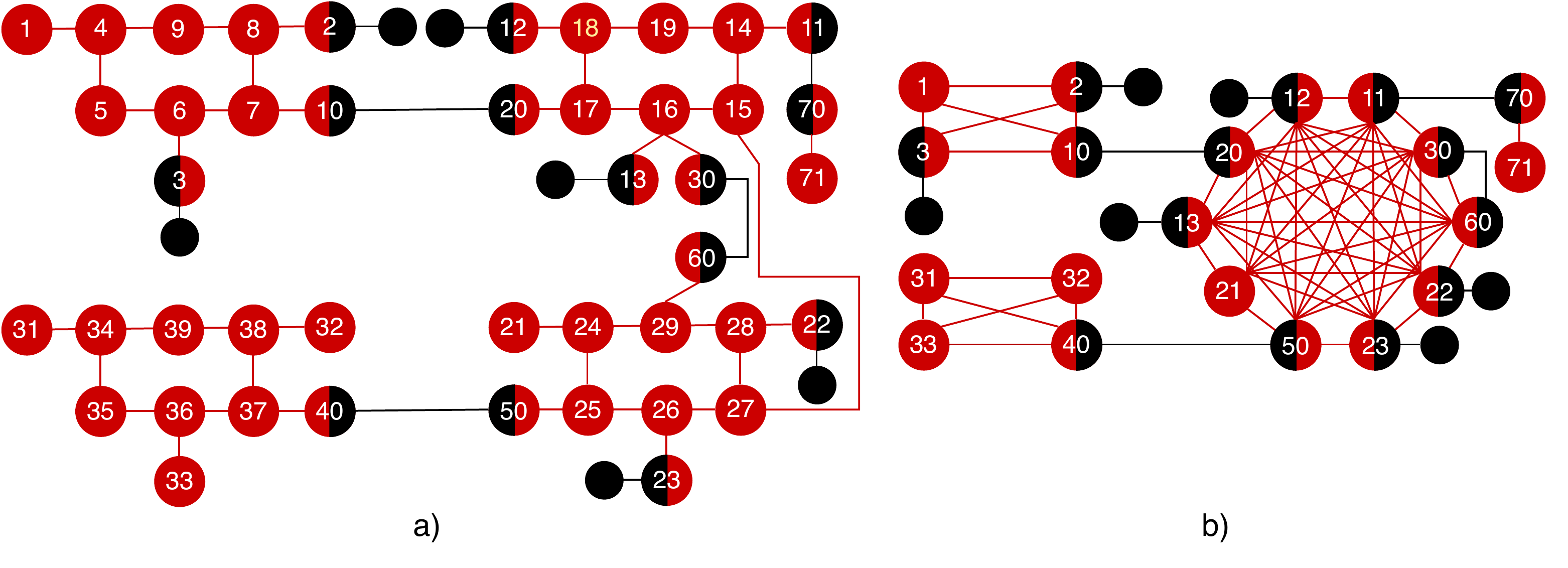} 
	\caption{a) Graph representation of the system in Fig.~\ref{fig:grid} and b) Kron-reduced graph obtained by removing interior nodes.	\label{fig:graph.grid}}
\end{figure}

For the purpose of stability analysis, we consider constant power loads that are commonly used in frequency stability analysis. This allows for Kron reduction of the ac and dc networks\footnote{While our results do not require Kron reduction of dc networks, we consider Kron reduced dc networks for simplicity of the notation.}. In other words, loads at nodes without   synchronous machine, converter, or dc power sources are mapped to nodes with machines, converters, or dc power sources. The original graph of the system in Fig.~\ref{fig:grid} is illustrated in Fig.~\ref{fig:graph.grid}~a) while Fig.~\ref{fig:graph.grid}~b) illustrates the Kron reduced graph.

Energy conversion devices are grouped into synchronous machines $\mc N_\ac$, dc/ac VSCs $\mc N_\cc$, and dc nodes $\mc N_\dc$. To each synchronous machine node $l\in \mc N_\ac$ we assign a voltage phase angle $\theta_l\in \mathbb{R}$, voltage magnitude $V_l\in \mathbb{R}_{\geq 0}$, and speed $\omega_l\in \mathbb{R}_{\geq 0}$. VSC nodes $l\in \mc N_\cc$ interconnect ac and dc networks (see Fig.~\ref{fig:graph.grid}) and are assigned a voltage phase angle $\theta_l\in \mathbb{R}$ and frequency $\omega_l\in \mathbb{R}_{\geq 0}$, voltage magnitude $V_l\in \mathbb{R}_{\geq 0}$, and  dc voltage $v_l\in \mathbb{R}_{\geq 0}$. Finally, to each dc node $l\in \mc N_\dc$ we assign the dc voltage $v_l\in \mathbb{R}_{\geq 0}$. Next, we distinguish between ac edges $\mc E_\ac \subseteq (\mc N_\ac \cup \mc N_\cc) \times (\mc N_\ac \cup \mc N_\cc)$ that model ac power flows and dc edges $\mc E_\dc \subseteq \mc (\mc N_\dc \cup \mc N_\cc) \times (\mc N_\dc \cup \mc N_\cc)$ that model dc power flows. The resulting connected, undirected graph, is defined as $\mc G_N\coloneqq (\mc N_N, \mc E_N)$, where $\mc N_N \coloneqq \mc N_\ac \cup \mc N_\cc \cup \mc N_\dc$ and $\mc E_N \coloneqq \mc E_\ac \cup \mc E_\dc$, and illustrated in Fig.~\ref{fig:graph.grid}~b).

For each node $l\in \mc N_\ac \cup \mc N_\cc$ we linearize the  quasi-steady-state ac power injection $P_{\ac,l} \in \mathbb{R}$ at $V_l^\star=1$~p.u. and $\theta_l^\star=\theta_k^\star$.  For each node $l \in \mc N_\dc \cup \mc N_\cc$ we linearize the quasi-steady-state dc power injection $P_{\dc,l} \in \mathbb{R}$ at $v_l^\star=1$~p.u to obtain 
\begin{subequations}
\begin{align}
	&\!P_{\delta,\ac,l}\!=\!\sum\nolimits_{(k,l)\in \mc E_\ac} \!\! {b}_{kl}^\ac(\theta_{\delta,l}\!-\!\theta_{\delta,k}) \!+ \!P_{d_\ac,l},\! \forall l\!\in\! \mc N_\ac \!\cup \mc N_\cc, \label{eq:ac.pf} \\
	&\!P_{\delta,\dc,l}\!=\!\sum\nolimits_{(k,l) \in \mc E_\dc} \! \! {g}_{kl}^\dc(v_{\delta,l}\!-\!v_{\delta,k})\! + \!P_{d_\dc,l},\! \forall l\!\in\!\mc N_\dc\! \cup \mc N_\cc. \label{eq:dc.pf}
\end{align}
\end{subequations}
Here ${b}_{kl}^\ac \in \R_{\geq 0}$, ${g}_{kl}^\dc \in \R_{\geq 0}$, $P_{d_\ac,l}\in \R$, and $P_{d_\dc,l}\in \R$  denote ac line susceptances, dc line conductances, and load perturbations mapped to the nodes of the Kron reduced graph.
\subsection{Power conversion}\label{sec:conversion.models}
The node sets $\mc N_\ac$ and $\mc N_\cc$ model power conversion devices, i.e., synchronous machines and power converters.
\subsubsection{Synchronous machine (SM)} 
\begin{figure}[t!]
	\centering
	\includegraphics[trim=6.5mm 1mm 1.05cm 0.3mm, clip, width=1\columnwidth]{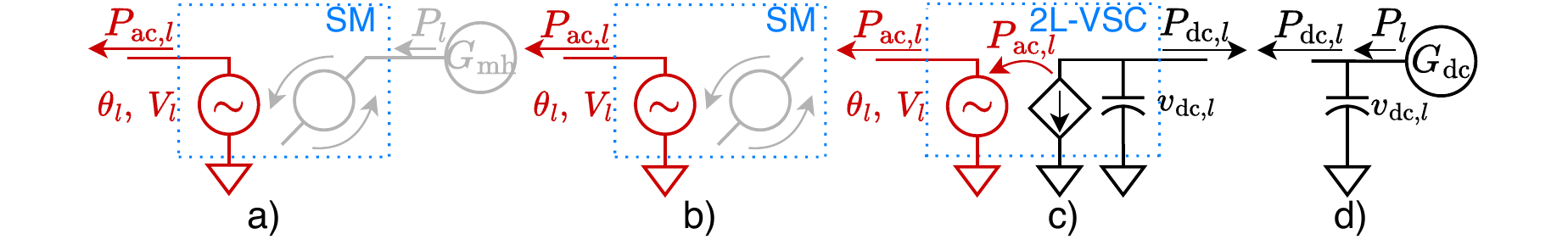}
	\caption{Power injections and bus voltages of a) a SM with a mechanical power source, b) a SM without a mechanical source (i.e., only rotating mass), c) a two-level VSC, and d) a dc bus with a dc power source.} \label{fig:poweer.directions}
\end{figure}

are shown in Fig.~\ref{fig:poweer.directions}~a) and modeled using the standard swing equation model
	\begin{align}\label{eq:SM.model} 
		\ddt \theta_{\delta,l} = \omega_{\delta,l}, 
		\quad J_l \omega_{l}^\star \ddt \omega_{\delta,l} = P_{\delta,l}-P_{\delta,\ac,l}, \ l \in \mc N_{\ac}, 
	\end{align}
with speed $\omega_{\delta,l}\in \R_{>0}$, machine inertia $J_l\in \R_{>0}$, and mechanical power applied to the rotor $P_{\delta,l}\in \R$. Notably, an SM that is not connected to mechanical power generation (i.e.,  $P_{\delta,l}=0$) can be used to model an energy storage element (e.g., flywheel) or synchronous condenser (SC) as shown in Fig.~\ref{fig:poweer.directions}~b). Broadly speaking, these devices respond to power imbalances during transients but cannot provide a sustained primary frequency control response.
\subsubsection{dc/ac VSC} 
For the purpose of analysis, we consider a lossless averaged model of a two-level VSC\footnote{Our control and analysis are applicable to more complex converter topologies, such as modular multilevel converters (MMCs)~\cite{GSP+21}.} illustrated in Fig.~\ref{fig:poweer.directions}~c). Using standard timescale separation arguments, the VSC output filter dynamics and inner control loops that track references for the ac voltage magnitude $V_{\delta,l} \in \mathbb{R}_{\geq0}$ and phase angle $\theta_{\delta,l}\in \mathbb{R}$ are neglected \cite{SG+20}. Thus, $V_{\delta,l}$ and $\theta_{\delta,l}$ are the remaining control inputs used by the GFM control introduced in Sec.~\ref{sec:control} and we obtain the dc-link capacitor charge dynamics
\begin{align} \label{eq:converter.model}
		C_l v_{l}^\star \ddt v_{\delta,l} &=- P_{\delta,\ac,l} - P_{\delta,\dc,l}, \ l \in \mc N_{\cc},
	\end{align}
with capacitance $C_l\in \R_{>0}$, dc power $P_{\delta,\dc,l}$ flowing into the dc network, and ac power $P_{\delta,\ac,l}$ flowing into the ac network. 
\subsection{Power sources}\label{sec:generation.models}
Finally, we abstractly model common power generation technologies through mechanical power sources (e.g., steam turbine, wind turbine) and dc power sources (e.g., PV).
\subsubsection{Turbine/governor} The prevalent first-order model 
\begin{align} \label{eq:turbine.gov.model}
	T_{\g,l} \ddt P_{\delta,l}=-P_{\delta,l} - k_{\g,l} \omega_{\delta,l}, 
\end{align}  
with turbine time constant $T_{\g,l} \! \in \! \R_{\geq 0}$, governor gain (i.e., sensitivity to frequency) $k_{\g,l} \! \in \! \R_{\geq 0}$, and turbine power $P_{\delta,l} \! \in \! \R$ is used to model legacy generation as shown in Fig.~\ref{fig:poweer.directions}~a). If the source generates constant power, it is modeled by $k_{\g,l}=0$.
\subsubsection{Controllable dc power sources} We model controllable dc power sources as power injection $P_{\delta,l} \! \in \! \R$, $l \!\in\! \mc N_\dc$ into dc nodes as shown in Fig.~\ref{fig:poweer.directions}~d). This results in the linearized dc bus dynamics
\begin{align}
	C_l v_{l}^\star \ddt v_{\delta,l} = P_{\delta,l} - P_{\delta,\dc,l},  l \in \mc N_\dc. \label{eq:pure.dc.node}
\end{align}
Within their limits, and on the time scales of interest, controllable DC sources (e.g., battery storage or two-stage PV system) can be modeled by 
\begin{align}	\label{eq:control.dc.sources}
	T_{\g,l} \ddt P_{\delta,l}=-P_{\delta,l}-k_{\g,l}v_{\delta,l},
\end{align} 
with time constant $T_{\g,l}\in \R_{> 0}$ and sensitivity $k_{\g,l}\in \R_{\geq 0}$ to dc voltage deviations. DC sources that generate constant power or track their MPP are modeled by $k_{\g,l}=0$.
\subsubsection{Wind turbine (WT)} A WT generates mechanical power that is converted to electrical power by a permanent magnet synchronous generator (PMSG) modeled via \eqref{eq:SM.model}. To obtain a linearized model of the WT mechanical power $P_{\delta,l}$, we first review the nonlinear wind turbine model (cf.~\cite{LSG22})
		\begin{align} \label{eq:wt.non.lin}
			T_{\g,l} \ddt \beta_l = -\beta_l+u_{l}^\beta, 
			\quad P_l =\tfrac{1}{2} \rho_l \pi R_l^2 C_{p,l}(\lambda_l, \beta_l) v_{\text{w},l}^3,
		\end{align}
with air density $\rho_l \in \R_{>0}$, rotor radius $R_l\in \R_{>0}$, wind speed $v_{\text{w},l}\in \R_{>0}$, tip speed ratio $\lambda_l=R_l\omega_{l}/v_{\text{w},l}$, pitch angle $\beta_l\in \R_{\geq0}$, time constant and control input of the pitch motor $T_{\g,l}\in \R_{> 0}$ and $u^\beta_{l}\in \R$. The function, $C_{p,l}\in \R_{>0}\times \R_{>0}\rightarrow \R_{>0}$ models the captured wind power and implicitly depends on the wind speed and rotor speed through the tip speed ratio. We assume the wind speed to be constant on the timescale considered in this manuscript. Extensions to time-varying wind speed are seen as an interesting topic for future work.  Figure~\ref{fig:operating_points} illustrates the WT power generation as a function of rotor speed $\omega_l$. To analyze \eqref{eq:wt.non.lin} in our framework, we linearize $P_l$ around a nominal operating point $(v_\w^\star,\omega_{l}^\star,\beta_l^\star)$ to obtain $P_{\delta,l} = -k_{\w,l} \omega_{\delta,l} - k_{\beta,l} \beta_{\delta,l}$ with sensitivities 
\begin{subequations} \label{eq:wt.lin.const}
	\begin{align}
		k_{\w,l} &\coloneqq -\frac{\partial  P_l}{\partial \omega_l}\big\vert_{(\omega_l,v_{\w,l},\beta_l)=(\omega^\star_l,v_{\w,l}^\star,\beta^\star_l)}	\in \mathbb{R}_{\geq 0}, \label{eq:wt.lin.const.w}\\
		k_{\beta,l} &\coloneqq -\frac{\partial  P_l}{\partial \beta_l}\big\vert_{(\omega_l,v_{\w,l},\beta_l)=(\omega^\star_l,v_{\w,l}^\star,\beta^\star_l)}  \in \mathbb{R}_{\geq 0},
	\end{align}
\end{subequations}
to changes in rotor speed and blade pitch angle. Combining $P_{\delta,l}$ with the (linear) pitch angle dynamics and applying the change of coordinates $P_{\delta,\beta,l} = k_{\beta,l} \beta_{\delta,l}$ we obtain
\begin{align} \label{eq:wt.lin.model}
\!\!\!\!\!\!\!	P_{\delta,l} \!=\! -k_{\w,l} \omega_{\delta,l}\! +\! P_{\delta,\beta,l},\ 
T_{\g,l} \ddt P_{\delta,\beta,l} \!=\! -P_{\delta,\beta,l} + k_{\beta,l} u^\beta_{\delta,l}.\!\!\!\!\!\! 
\end{align}
\subsubsection{Solar photovoltaics (PV)} 
are connected to the network through dc nodes \eqref{eq:pure.dc.node}. To obtain a linearized model of the dc power $P_{\delta,l}$ generated by a PV panel, we first introduce the single-diode model of the PV current $i_l\in \R$~\cite[Fig. 4]{GRR2009}
\begin{align} \label{eq:PV.model}
	i_l\!=\!i_{\text{L},l}-i_{0,l}\big(\text{exp}({\tfrac{v_l+R_{s,l}i_l}{v_{t,l}\alpha_l}})-1\big)-\tfrac{v_l+R_{s,l}i_l}{R_{p,l}},
\end{align}
where $v_{l}\in \R_{\geq 0}$, $i_{\text{L},l}\in \R_{\geq 0}$, and $i_{0,l}\in  \R_{\geq 0}$ denote the dc voltage, photovoltaic current, and saturation current. The thermal voltage of the PV's cell array is $v_{t,l}\in \R_{>0}$, while $R_{s,l}\in \R_{>0}$ and $R_{p,l}\in \R_{>0}$ are series and parallel resistances. Fig.~\ref{fig:operating_points} illustrates the PV power as a function of dc voltage.

For analysis purposes, we linearize $P_l=v_l i_l$ at the desired operating point (i.e., dc voltage) and obtain $P_{\delta,l}=-k_{\pv,l}v_{\delta,l}$ with sensitivity $k_{\pv,l}\coloneqq - \frac{dP_l}{dv_l}\vert_{v_l=v_l^\star} \in \R_{\geq 0}$ to dc voltage deviations.
In the remainder of the manuscript, we assume that the operating point of  renewable generation is periodically updated (see, e.g., \cite{HS+17} and \cite{LSG22} for details) and changes slowly compared to the time-scales of interest.

\subsection{Curtailment \& local control of renewables} \label{sec:op.point}
Independently of the converter control strategy used, the ability of renewable generation to provide grid support strongly depends on its operating point and curtailment strategy. 
In particular, only a renewable source operated below its MPP (i.e., $P_l^\star<P_l^\mpp$), can adjust its power output up or down to contribute to frequency/dc voltage stabilization~\cite{SG21}. 
\subsubsection{Solar PV}
Curtailment of PV is achieved by operating at a dc voltage above the MPP voltage, i.e., $v_l^\star>v_l^\mpp$ (see Fig.~\ref{fig:operating_points}). At this operating point, the PV sensitivity $k_\pv$  corresponds to the slope of the curve at the operating point $(P_l^\star, v_l^\star)$. Thus, if $v_l^\star>v_l^\mpp$, then $k_{\pv,l}\in\R_{>0}$, and if $v_l^\star=v_l^\mpp$, $k_{\pv,l}=0$. In other words, if $v_l^\star>v_l^\mpp$, PV provides proportional dc voltage control. In contrast, at the MPP (i.e., $v_l^\star=v_l^\mpp$), the PV does not respond to dc voltage deviations. Moreover, if $v_l^\star<v_l^\mpp$, the physics of PV are inherently unstable (i.e., $k_{\pv,l}\in\R_{<0}$) \cite{HS+17}. Because there is no significant advantage to operating in the unstable region, we only consider operating points in the stable region.
\subsubsection{Wind generation}
\begin{figure}[t!]
	\centering
	\includegraphics[trim=2mm 1.5mm 1.15mm 1.6mm, clip, width=1\columnwidth]{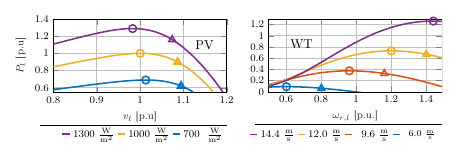}
	\caption{Power generation of PV (left) as a function of  dc voltage and irradiation (constant temperature of $25^\circ$~C) and WT (right) as a function of rotor speed and wind speed  (zero blade pitch angle). The MPP and a (stable) operating point at 90\% MPP are denoted by circles and triangles. \label{fig:operating_points}}
\end{figure}

Curtailment of WTs can be achieved by increasing the rotor speed beyond its optimal speed $\omega_{l}^\mpp$ (i.e., $\omega_{l}^\star > \omega_{l}^\mpp$, see Fig.~\ref{fig:operating_points}) and/or by increasing the blade pitch angle (i.e., $\beta_l^\star\in\R_{>0}$) (cf. \cite{LSG22,AFP16}). In particular, rotor speed-based curtailment (i.e., $\omega_{l}^\star > \omega_{l}^\mpp$) results in $k_{\w,l}\in\R_{>0}$ and can be interpreted as proportional WT speed control. While our analytical framework allows modeling speed-based and pitch angle-based curtailment, our case studies will prioritize rotor speed-based curtailment over pitch angle-based curtailment to leverage the significant kinetic energy storage of WTs using speed-based curtailment. Finally, we assume that $\omega_{l}^\star \geq \omega_{l}^\mpp$ to exclude the inherently unstable operating region $\omega_{l}^\star < \omega_{l}^\mpp$, i.e., $k_{\w,l}\in\R_{<0}$ \cite{LSG22}. 

While commercial WTs often rely on proportional-integral (PI) control of the rotor speed through the pitch angle for MPPT control, we use the proportional rotor speed control $u^\beta_l\coloneqq \beta_l^\star - k_{\pb,l}(\omega_{l}-\omega_l^\star)$ with control gain $k_{\pb,l}\in \R_{\geq 0}$ to mitigate rotor speed deviations (i.e., power imbalances) and enable grid-support functions. Substituting $u_{\delta,l}^\beta=u_l^\beta-\beta^\star_l$ into \eqref{eq:wt.non.lin}, we can express the pitch angle-based response in the form of the turbine/governor system \eqref{eq:turbine.gov.model} with WT pitch angle power generation sensitivity $k_{\g,l}\coloneqq k_{\beta,l}k_{\pb,l}\in \R_{\geq 0}$.

\subsection{Modeling complex power systems}  \label{sec:illustrative example}
\begin{figure}
	\includegraphics[ width=\columnwidth]{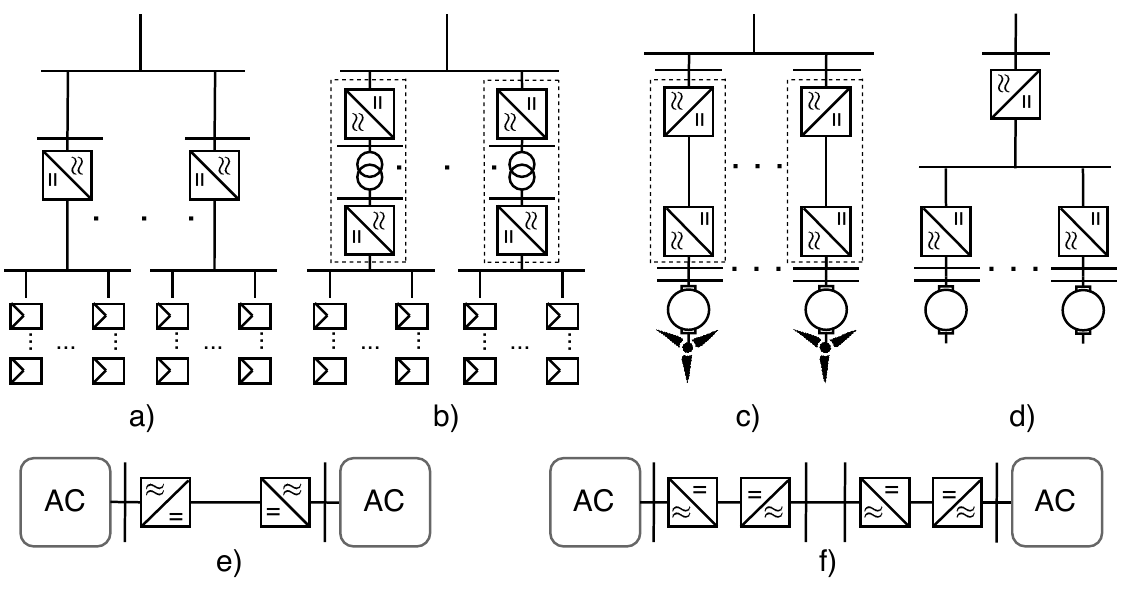} 
	\caption{a) PV plant and dc/ac power converters b) PV plant dc collector network with dual active bridge converter, c) wind farm with PMSG WTs d) flywheel energy storage system and with dc network e) high voltage dc (HVDC) link, and f) low frequency ac (LFAC) connection. \label{fig:sources}}
\end{figure}

A wide range of complex emerging systems (e.g., Fig.~\ref{fig:grid} and Fig.~\ref{fig:sources}), can be readily modeled in our framework through composition of device models (see Sec.~\ref{sec:power.network.topology}-Sec.~\ref{sec:generation.models}). For instance, fuel-based legacy SGs are modeled as a combination of a synchronous machine and turbine/governor system. PV plants with standard single-stage solar PV systems are modeled as interconnection of a dc/ac VSC and a PV module through a dc edge as shown in Fig.~\ref{fig:sources}~a) while advanced configurations with dc collector networks and, e.g., dual active bridge (DAB) converters \cite{BD18} can be modeled by connecting ac terminals of dc/ac VSCs as in Fig.~\ref{fig:sources}~b). PMSG WTs are modeled by combining two dc/ac VSCs with an SM and the WT aerodynamics as shown in Fig.~\ref{fig:sources}~c). Note that devices consisting of the multiple converters and/or machines (e.g., PMSG WT) are modeled using multiple buses and the bus index $l$ is used to differentiate individual components (e.g., see  buses~70 and~71 in Fig.~\ref{fig:grid}). Finally, emerging transmission technologies can be readily modeled. For example, high voltage dc (HVDC) transmission \cite{GSP+21} is modeled by connecting two dc/ac VSCs via a dc line as in Fig.~\ref{fig:sources}~e) and low-frequency ac (LFAC) transmission \cite{SR22} is modeled by four ac/dc power converters as in Fig.~\ref{fig:sources}~f). By combining these models, one can, e.g., readily model an offshore wind farm as a combination of PMSG WTs and a point-to-point HVDC link. Large-scale flywheel storage system that connect multiple flywheels through a dc collector network can be modeled using SMs (without generation, $P_{\delta,l}=0$), dc/ac VSCs, and a dc network as in Fig.~\ref{fig:sources}~d).
\section{Universal dual-port GFM control} \label{sec:control}
\begin{figure}[t]
	\centering
	\includegraphics[ width=0.5\textwidth]{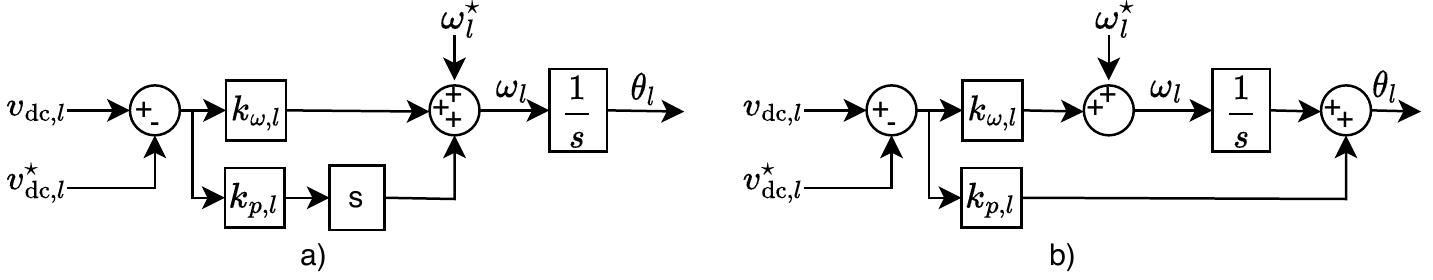} 
	\caption{Block diagram of a) a PD realization~\eqref{eq:control.law} and b) a PI realization~\eqref{eq:control.pi.interpretation} of dual-port GFM control.	\label{fig:block_diag}}
\end{figure}

The ac voltage phase angle dynamics of each dc/ac VSC, are prescribed by the universal dual-port GFM control, i.e., the proportional-derivative (PD) $v_{\delta}-f$ droop
\begin{align} \label{eq:control.law}
	\ddt \theta_{\delta,l} =\omega_l-\omega_l^\star =  k_{p,l} \ddt v_{\delta,l} + k_{\omega,l} v_{\delta,l}
\end{align}
with proportional and derivative gains $k_{\omega,l} \in \R_{>0}$ and $k_{p,l} \in \R_{>0}$. The controller resembles machine emulation control~\cite{CBB+2015,MDPS+17}, but adds a dc voltage setpoint (i.e., $v_{\delta,l}\!=\!v_l\!-\!v^\star_l$) and derivative feedback see Fig.~\ref{fig:block_diag}~a). Moreover, the lead-lag control of the squared dc voltage proposed in~\cite{HHH+18} can be understood as an approximation of the PD control \eqref{eq:control.law}. We emphasize that \cite{CBB+2015,MDPS+17,HHH+18} all assume a controllable dc source and do not analyze the dynamics of renewables, HVDC, energy storage devices, and interactions of the aforementioned technologies with legacy power systems (e.g., SGs and SCs).

Thus, while variants of \eqref{eq:control.law} have been proposed before, they have only been analyzed in the context of machine emulation and, to the best of the authors' knowledge, no  end-to-end stability results that account for the dynamics of (renewable) generation are available. Crucially, universal dual-port GFM control propagates power imbalances to power sources and makes the power generation (e.g., PV, wind) transparent in the stability analysis. In this context, we observe that \eqref{eq:control.law} can perform either approximate MPPT or provide common grid support functions depending on the operating point of renewables (see Sec.~\ref{sec:op.point}). Specifically, depending on the capabilities of the power source and its operating point either (i) only oscillation damping, (ii) only a fast frequency response, or (iii) only an inertia response, or (iv) both a fast frequency response and inertia response are provided. Moreover, the derivative feedback in \eqref{eq:control.law} is crucial to (i) tune the transient response, (ii) to ensure stability of network circuit dynamics~\cite{G22}, and (iii) for our stability analysis in Sec.~\ref{sec:analysis}. FFinally, similar to power-balancing dual-port GFM control~\cite{SG21}, the controller exhibits bidirectional grid support and unifies GFL and GFM functions in a single control without mode switching. The reader is referred to Sec.~\ref{sec:comparison} for a detailed comparison of \eqref{eq:control.law} and the control proposed in~\cite{SG21}.
\subsection{Implementation of universal dual-port GFM control}
Next, we seek a realizable implementation of \eqref{eq:control.law} that avoids the ideal differentiator. In this context, \cite{HHH+18} can be understood as a lead-lag approximation of \eqref{eq:control.law} that introduces additional dynamics in the system and may lead to undesirable responses and amplify measurement noise and switching ripple. Instead, we leverage the fact that the inner converter control requires an ac voltage phase angle and magnitude reference not the frequency. Integrating the frequency in \eqref{eq:control.law} (i.e., $\theta_l=\int\nolimits_{t_0}^t \omega_l \Delta \tau$) results in a derivative-free implementation of \eqref{eq:control.law} using the proportional-integral (PI) control%
\begin{align}
	\theta_{l}\!= \!\theta_l^\star\!+\!\Delta \theta_l+ k_{p,l} v_{\delta,l} + k_{\omega,l} \gamma_l, \quad \ddt \gamma_l = v_{\delta,l}. \label{eq:control.pi.interpretation}
\end{align}
shown in Fig.~\ref{fig:block_diag}~b). In the remainder, we use \eqref{eq:control.law} for analysis and theoretical results  while the derivative-free implementation is used in our case studies. We note that the derivative-free implementation does not require knowledge of an absolute phase angle $\theta_l$ or phase angle setpoint $\theta^\star_l$. In particular, the lack of a global reference angle can be modeled as a constant offset $\Delta \theta_l$ to the phase angle that is then eliminated by the integral part of the PI controller. Notably, $\Delta \theta_l$ does not impact stability analysis and analytical results and does not appear in \eqref{eq:control.law} because $\ddt \Delta \theta_l = 0$.

\subsection{Comparison with power-balancing dual-port GFM control} \label{sec:comparison}
The control \eqref{eq:control.law} exhibits similar properties to power-balancing dual-port GFM control \cite{SG21}
\begin{align} \label{eq:control.law.pbdp}
	\ddt \theta_{\delta,l} = \omega_l-\omega_l^\star =  - m_{p,l} P_{\delta,\ac,l} + k_{\omega,l} v_{\delta,l}
\end{align}
with droop coefficient $m_{p,l}\in \mathbb{R}_{>0}$. In particular, under the lossless VSC model \eqref{eq:converter.model}, shown in Fig.~\ref{fig:poweer.directions}~c), \eqref{eq:control.law} becomes
\begin{align} \label{eq:control.law.lossless}
	\ddt \theta_{\delta,l} = \omega_l-\omega_l^\star =  - \tfrac{k_{p,l}}{C_l v_l^\star} (P_{\delta,\ac,l}+P_{\delta,\dc,l}) + k_{\omega,l} v_{\delta,l}
\end{align}
In other words, \eqref{eq:control.law} does not require a power setpoint and instead acts on the difference between ac and dc power through the $\ddt v_{\delta,l}-f$ droop term. While it may seem counter-intuitive to omit $P_\ac\!-\!f$ droop from the control, we highlight that the objective is to stabilize the VSC's ac and dc terminal voltages and balance the system, not to operate close to a prescribed power transfer between dc and ac terminals. In particular, the $P_\ac\!-\!f$ and $v_{\delta}\!-\!f$ droop terms in \eqref{eq:control.law.pbdp} represent conflicting objectives (i.e., operating close to a power setpoint vs. balancing ac and dc systems) and counteract each other. This may result in significant oscillations during transients and counter-intuitive responses when post-contingency power flows significantly differ from the VSC power set-point \cite{GSP+21}.

Instead, \eqref{eq:control.law}, maps signals indicating power imbalance
(i.e., ac frequency and dc voltage deviation) between the VSC terminals without requiring measurements or estimates of the ac grid frequency. In other words, by decreasing or increasing the ac frequency and dc voltage, \eqref{eq:control.law} passes on power imbalances to power generation in the ac and dc networks that respond according to their sensitivities established in Sec.~\ref{sec:modeling}. 

Next, we show that the overall system is stable under mild assumptions on the network topology and number of devices with non-zero sensitivities. Additionally, in contrast to \eqref{eq:control.law.pbdp}, the steady-state response of \eqref{eq:control.law} is uniquely characterized by $k_{\omega,l}$  and $k_{p,l}$ can be used to adjust the transient response. The focus of this work are stability conditions that ensure stability for a wide range device and of topologies. Control tuning within the set of stabilizing gains to enhance performance is seen as an interesting topic for future work.
\begin{table*}[b!!]
	\begin{align}
	\!\!\!\!\!\!	\resizebox{2\columnwidth}{!}{
		\setlength\tabcolsep{0.9pt}
		\bgroup
		\def\arraystretch{1}%
	\label{eq:matrix.A}
	$	T\! \ddt \!\! \begin{bmatrix}
			\eta_\delta \\ \omega_\delta \\v_\delta \\ P_{\fr,\delta} \\ P_{\zs,\delta}
		\end{bmatrix} \!\! \!= \! \!\! \underbrace{\begin{bmatrix}
			B_\eta \mc I_{\da}^\T \mc I_{\ad} B_\ac \mc W_\ac  &  (\mc I_{\ac} B_\ac)^\T  & (\mc I_{\ad} B_\ac)^\T K_\omega  \mc I_{\da}\! -B_\eta L_\dc & \mathbbl{0}_{|\mc E_{\ac}| \times |\mc N_\fr|} & \mathbbl{0}_{|\mc E_\ac|\times |\mc N_\zs|}\\
			-\mc I_{\ac} B_\ac \mc W_\ac & -\mc I_\w^\T K_\w \mc I_\w & \mathbbl{0}_{|\mc N_\ac| \times |\mc N_\cc \cup \mc N_\dc|} & \mc I_{\fr,\ac}^\T & \mc I_{\zs,\ac}^\T\\
			-\mc I_{\da}^\T \mc I_{\ad} B_\ac \mc W_\ac  & \mathbbl{0}_{|\mc  N_\cc \cup \mc N_\ac|\times |\mc N_\ac|}  & - L_\dc-(\mc I_{\pv}  \mc I_{\dc} )^\T K_\pv \mc I_{\pv} \mc I_\dc & (\mc I_{\fr,\dc} \mc I_\dc) ^\T & ( \mc I_{\zs,\dc} \mc I_\dc)^\T \\
			\mathbbl{0}_{|\mc N_\fr |\times |\mc E_\ac|} & - K_\g\mc I_{\fr,\ac}& - K_\g \mc I_{\fr,\dc} \mc I_{\dc} & -I_{|\mc N_\fr|} & \mathbbl{0}_{|\mc N_\fr| \times |\mc N_\zs|} \\
			\mathbbl{0}_{|\mc N_\zs | \times |\mc E_\ac|}  & \mathbbl{0}_{|\mc N_\zs  |\times |\mc N_\ac|}  & \mathbbl{0}_{|\mc N_\zs |\times |\mc N_\cc \cup \mc N_\dc|}   & \mathbbl{0}_{|\mc N_\zs|\times |\mc N_{\fr}| } & -I_{|\mc N_\zs|}
		\end{bmatrix}}_{\eqqcolon A} \!\! \! \begin{bmatrix}
			\eta_\delta \\ \omega_\delta \\v_\delta \\ P_{\fr,\delta} \\ P_{\zs,\delta}
		\end{bmatrix} \! \! + \! \! \underbrace{\left[\begin{array}{@{}cc@{}} \;\,B_\eta & B_\eta\!\!\!\! \\ \hline 
			\multicolumn{2}{c}{\multirow{2}{*}{$-I_{n_d}$}}\\
			\\ \hline 
			\multicolumn{2}{c}{\multirow{2}{*}{$\!\!\!\!\mathbbl{0}_{|\mc N_{\fr} \cup \mc N_{\zs}| \times n_d}\!\!\!\!$}}\\
			\\
		\end{array}\right]}_{\eqqcolon B} \! \! P_d$
	\egroup	}\!\!\!
\end{align}
\end{table*}
\section{End-to-end Stability and Steady-State analysis}\label{sec:analysis}
To systematically model and analyze the system, we start from the Kron reduced graph $\mc G_N$ and distinguish between the ac graph $\mc G_\ac\coloneqq (\mc N_\ac \cup \mc N_\cc,\mc E_\ac)$ consisting of the nodes and edges associated with ac quantities, and the dc graph $\mc G_\dc\coloneqq (\mc N_\dc \cup \mc N_\cc,\mc E_\dc)$ consisting of the nodes and edges associated with dc quantities, i.e., $\mc G_N= \mc G_\ac \cup \mc N_\dc$. Moreover, we stress that converter nodes are associated with both ac and dc quantities (see Fig.~\ref{fig:poweer.directions}~c)). Hence, converter nodes are part of both ac and dc subnetworks, i.e., $\mc G_\ac \cap \mc G_\dc =\mc N_\cc$.

\begin{figure}[t]
	\centering
	\includegraphics[trim=0 2mm 0 0, clip, width=0.5\textwidth]{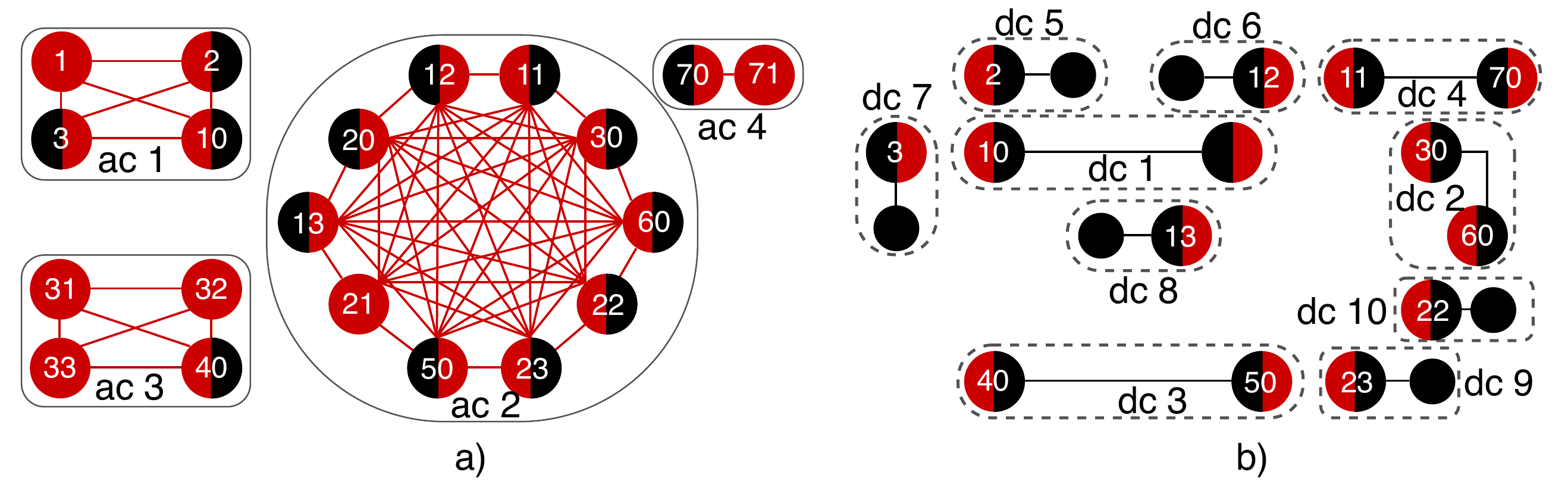} 
	\caption{Illustration of the a) connected ac subnetworks and b) connected dc subnetworks of the system in Fig.~\ref{fig:grid}.\label{fig:subnetworks}}
\end{figure}

The ac subnetwork $\mc G_\ac=\bigcup_{i=1}^{N_\ac} \mc G_\ac^i$ is formally defined via $N_\ac$ maximal connected ac subnetworks $\mc G_\ac^i \coloneqq (\mc N_\ac^i \cup \mc N_{\ad}^i, \mc E_\ac^i)$, where $\mc E_\ac^i$ is edge set and $\mc N^i_\ac \subseteq \mc N_\ac$, $\mc N_\ad^i \subseteq \mc N_\cc$ are ac and dc/ac nodes. Note that a maximal connected ac subnetwork is defined as a maximal connected (i.e., non-overlapping) component consisting only of ac nodes and edges. Analogously, the dc subnetwork $\mc G_\dc \!=\!\bigcup_{i=1}^{N_\dc} \mc G_\dc^i$ consists of the $N_\dc$ maximal connected dc subnetworks $\mc G_\dc^i \coloneqq (\mc N_\dc^i 
\! \cup \! \mc N_{\da}^i, \mc E_\dc^i)$ with edge set $\mc E_\dc^i$, dc nodes $\mc N^i_\dc \!\subseteq \!\mc N_\dc$, and dc/ac nodes $\mc N_\da^i\! \subseteq\! \mc N_\cc$. Figure~\ref{fig:subnetworks}~b) illustrates partitioning of the system shown in Fig.~\ref{fig:grid} into maximally connected ac and dc subnetworks.

To facilitate our stability analysis, we distinguish between five broad sets of nodes within  the set of the ac and dc nodes  ($\mc N_\ac \cup \mc N_\dc$): i) ac nodes $\mc N_\fr\subseteq \mc N_\ac \cup \mc N_\dc$ connected to power generation that responds to frequency/dc voltage deviations (i.e., \eqref{eq:turbine.gov.model}, \eqref{eq:control.dc.sources}, or \eqref{eq:wt.lin.model} with $k_{\g,l}>0$), ii) ac nodes $\mc N_\zs\subseteq \mc N_\ac \cup \mc N_\dc$ connected to power generation that does not respond to frequency/dc voltage deviations (i.e.,\eqref{eq:turbine.gov.model}, \eqref{eq:control.dc.sources}, or \eqref{eq:wt.lin.model} with $k_{\g,l}=0$), iii) dc nodes $\mc N_\pv\subseteq \mc N_\dc$ connected to curtailed PV (i.e., $k_{\pv,l}\!>\!0$), iv) ac nodes $\mc N_\w\!\subseteq\! \mc N_\ac$ connected to WTs using rotor speed-based curtailment (i.e., $k_{\w,l}\!>\!0$) and, v) the remaining machine nodes $\mc N_\ac \setminus (\mc N_\fr \cup \mc N_\zs \cup \mc N_\w)$, not connected to generation (e.g., SC or flywheel storage system).

\subsection{Overall linearized model}
The Vectors $\theta_\delta\in \R^{|\mc N_\ac \cup \mc N_\cc|}$, $\omega_\delta \in \R^{|\mc N_\ac \cup \mc N_\cc|}$ and $v_\delta \in \R^{|\mc N_\cc \cup \mc N_\dc|}$ collect phase angle, frequency, and dc voltage deviations at all nodes. Power injections of the sources providing frequency response (i.e., $k_{\g,l}>0$) are collected in $P_{\fr,\delta} \in \R^{|\mc N_\fr|}$. Similarly, power injections of the sources that do not provide a frequency response (i.e., $k_{\g,l}=0$) are collected in $P_{\zs,\delta} \in \R^{|\mc N_\zs|}$. To simplify the analysis, we change coordinates from phase angles $\theta_\delta$ to phase angle differences, i.e., $\eta_\delta \coloneqq B_\ac^\T \theta_\delta$ \cite[cf. Sec. III]{MDPS+17}, where $B_\ac \in \{ -1,0,1 \}^{|\mc N_\ac \cup \mc N_\ad| \times |\mc E_\ac|}$ is the oriented incidence matrix of the ac graph $\mc G_\ac$ with $|\mc N_\ac \! \cup \! \mc N_\cc|$ nodes and $|\mc E_\cc|$ edges. In other words, for a line $(l,k) \in \mc E_\ac$ enumerated by $m \in \{1,\ldots,|\mc E_\ac|\}$, the entry $\eta_{\delta,m}$ of the vector $\eta_{\delta}$ corresponds to the phase angle differences between two nodes connected via line $m$ relative to their phase angle differences at the nominal operating point, i.e., 
\begin{align*}
	\eta_{\delta,m} = \theta_{\delta,l}-\theta_{\delta,k}=\theta_l-\theta_k-(\theta_l^\star-\theta_k^\star)=\eta_m-\eta_m^\star.
\end{align*}
The Laplacian  $L_\dc$ corresponds to the dc graph $\mc G_\dc$ and the matrix $\mc W_\ac\coloneqq \{b^\ac_{lk}\}_{(l,k)\in \mc E_\ac}$ collects ac line susceptances $g^\ac_{lk}$ (see \cite{SG21}). The matrix $T\!\coloneqq \!\blkdiag \{ I_{|\mc N_\ac \cup \mc N_\cc|}, J, C, T_\fr,{T}_\zs \}$ collects i) machine inertia constants $J\coloneqq \diag \{ J_l \omega_l^\star \}_{l=1}^{|\mc N_\ac|} \succ 0$, ii) dc capacitances $C\coloneqq \diag \{ C_{l} v_l^\star \}_{l=1}^{|\mc N_\cc \cup \mc N_\dc|} \succ 0$, iii) time constants of the power generation providing frequency response $T_\fr \coloneqq \{ T_{\g,l} \}_{l=1}^{|\mc N_\fr|}\succ 0$, and iv) the time constants of the generation with zero sensitivity ${T}_\zs \coloneqq \{ T_{\g,l} \}_{l=1}^{|\mc N_\zs|} \succ 0$. Additionally, $K_\w\coloneqq \diag\{k_{\w,l}\}_{l=1}^{|\mc N_\w|}\succ 0$, $K_\pv \coloneqq \diag\{k_{\pv,l}\}_{l=1}^{|\mc N_\pv|}\succ 0$, and $K_{\g}=\diag\{k_{\g,l}\}_{l=1}^{|\mc N_\fr|} \succ 0$, collect the sensitivities of WTs using rotor speed-based curtailment, curtailed PV, and generation \eqref{eq:turbine.gov.model} or \eqref{eq:control.dc.sources} providing frequency or dc voltage control. Finally, the interconnection matrices $\mc I _{\ad}$, $\mc I_{\da}$, $\mc I_\ac$, $\mc I_\dc$, $\mc I_{\fr,\ac}$, $\mc I_{\fr,\dc}$, ${\mc I}_{\zs,\ac}$, ${\mc I}_{\zs,\dc}$, $\mc I_{\w}$ and $\mc I_\pv$ model the interconnection of devices to buses and are defined in~App.~\ref{app:intercon}. For instance, the entry $(i,j)$ of the interconnection matrix $\mc I_\w \in \{0,1\}^{|\mc N_\w| \times |\mc N_\ac|}$ is $1$ if the $j$th ac node is connected to the $i$th WT. Combining the models from Sec.~\ref{sec:modeling} and \eqref{eq:control.law}, the overall system model is given by \eqref{eq:matrix.A} where $P_d\coloneqq(P_{d_\ac},P_{d_\dc}) \in \R^{n_d}$, $n_d\coloneqq |\mc N_\ac| + |\mc N_\dc| + 2 |\mc N_\cc|$  $B_\eta \coloneqq -(\mc I_{\ad} B_\ac)^\T \!K_p\mc I_{\da} C^{-1}$. In the remainder, the load variations $P_{d}$ in the system \eqref{eq:matrix.A} are treated as constant disturbances.
\subsection{Stability conditions on the VSC control gains}\label{sec:stab.cond.and.discussion}
For simplicity, we assume that the VSC control gains $k_{\omega,l}$ are identical on every dc subnetwork $\mc G_\dc^i$, $i\in \N_{[1,N_\dc]}$.
\begin{condition}~\cite[Cond.~1]{SG21} \textbf{(Consistent $\boldsymbol{v}_\dc\!-\!\boldsymbol{f}$ droop)}
\label{assump:identical.k.theta.gains} For each $i \!\in\! \N_{[1,N_\dc]} $ and all $(n,l) \! \in\! \mc N^i_{\da} \! \times \! \mc N^i_{\da}$, it holds that $k_{\omega,n}\!=\! k_{\omega,l} \coloneqq k_\omega^i$.
\end{condition}
Condition~\ref{assump:identical.k.theta.gains} is required to enforce a consistent mapping between ac frequency deviations and dc voltage deviations across each maximal connected dc subnetwork but does not preclude asynchronous ac subnetworks with different nominal frequencies. For a detailed discussion see Sec~\ref{sec:cond1ss}. Condition~\ref{assump:identical.k.theta.gains} can be lifted for ac subnetworks that do not need to synchronize to the same frequency (see \cite[Thm. 1]{LSG22}) or two ac systems connected only by point-to-point HVDC. The following condition on the dc line resistance $r^{\dc}_{l,k} =1/g^{\dc}_{l,k}$, scaled dc bus capacitance $c_l$, and control gains $k_{p,l}$ and $k^i_\omega$ is used to establish stability of \eqref{eq:matrix.A} (see Sec.~\ref{sec:cond2transient} for an interpretation). To this end, we define the equivalent (parallel) dc resistance $r^{\dc}_{\text{eq},l}$ relative to node $l \in \mc N_{\da}^i$ as $\tfrac{1}{r^{\dc}_{\text{eq},l}} = \sum_{(l,k) \in  \mc E_\dc^i} \tfrac{1}{r^{\dc}_{lk}}$.
\begin{condition} \label{lemma:more.conservative.gains.assumption} \textbf{(Stabilizing control gains)} For each $i \! \in \! \N_{[1,N_\dc]}$ and all $l \! \in\! \mc N_{\da}^i$,  the gains $k_{p,l} \!\in \! \R_{>0}$ and $k^i_\omega \! \in \! \R_{>0}$ satisfy $k_{p,l} < 2 k^i_\omega c_l r^{\dc}_{\text{eq},l}$.
\end{condition}
\subsection{Stability conditions on the ac network topology}
We emphasize that, in the general setting considered in this work, the system is not asymptotically stable for all network topologies and line parameters~(cf. \cite[Ex.~1]{SG21}). Thus, in addition to Cond.~\ref{assump:identical.k.theta.gains} and Cond.~\ref{lemma:more.conservative.gains.assumption}, the system topology needs to be restricted. First, at least one power source is required to respond to frequency or dc voltage deviations.
\begin{condition}{\cite[Assump.~2]{SG21}} \label{assump:onestab} \textbf{(Frequency \& dc voltage stabilization)} 
	There exists $l \in \mc N_N$ such that $k_{\g,l}>0$ or $k_{\pv,l}>0$ or $k_{\w,l}>0$ (i.e., $\mc N_\fr \cup \mc N_\pv \cup \mc N_\w \neq \emptyset$).
\end{condition}
 While there are no restrictions on dc network topologies, we impose topological stability conditions that can be verified independently for each ac subnetwork $\mc G^i_\ac$ and do not require knowledge of the line parameters. Conditions for a wide range of ac subnetwork topologies and, e.g., $N\!-\!1$ stability, can be obtained by leveraging \cite[Alg.~1]{SG21}. Instead, in this work, we apply the following simplified condition that covers the vast majority of practically relevant topologies. The number of edges that are incident to the node $l$ is denoted with $\deg{(l)}$.
\begin{condition}\textbf{(Synchronizing ac connections)}
	\label{cond.syncac}
	Given node sets $\mc N_1$ and $\mc N_2 \!\supseteq\! \mc N_3$, we define the graph $\mc E_{\mc N_1,\mc N_2} \!\coloneqq\! \mc E^i_\ac \cap (\mc N_1 \!\times\! \mc N_2)$. For every $k \!\in\! \mc N_3$, there exists $l \in \mc N_1$ such that $(l,k) \in \mc E_{\mc N_1,\mc N_2}$ and $\deg{(l)}=1$.
\end{condition}

\begin{figure}[t]
	\centering
	\includegraphics[width=0.33\textwidth]{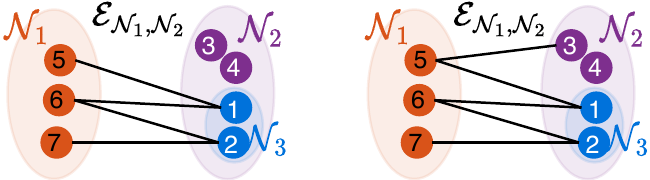} 
	\caption{Illustration of Cond.~\ref{cond.syncac}. The topology on the left satisfies Cond.~\ref{cond.syncac} because nodes 1 and 2 are each connected to a node with degree one (i.e., node 5 and 7). In contrast, the topology on the right does not satisfy Cond.~\ref{cond.syncac} since node 1 is not connected to a node with degree one.	\label{fig:syncac.example}}
\end{figure}

Topologies that do and do not satisfy Cond.~\ref{cond.syncac} are shown in Fig.~\ref{fig:syncac.example}.  To apply Cond.~\ref{cond.syncac} to the ac subnetworks $\mc G_\ac^i$, we distinguish between i) ac nodes $\mc N_{\ac^\fr}^i$ with frequency control (i.e., $k_{\g,l}>0$, $k_{\w,l}>0$) and ii) dc/ac nodes $\mc N_{\ad^\fr}^i$ connected to a dc network with at least one node providing dc voltage control (i.e., $k_{\g,l}>0$, $k_{\pv,l}>0$). Finally, $\mc N_{\ac^\oo}^i\! \coloneqq \mc N_\ac^i \!\setminus \mc N_{\ac^\fr}^i$ collects nodes corresponding to SCs and $\mc N_{\ad^\oo}^i \! \coloneqq \mc N_\ad^i \!\setminus\mc N_{\ad^\fr}^i$ collects the remaining ac/dc nodes (e.g., HVDC VSCs). 
\begin{condition} \label{cond:topology.both} \textbf{(Simplified conditions for synchronization)} Every ac subgraph $i \in \N_{[1,N_\ac]}$ satisfies Cond.~\ref{cond.syncac} with either
	\begin{enumerate}[label=\roman*)]
		\item \label{cond:few.machines.zs} $\mc N_1\!=\!\mc N_{\ac^\fr}^i\! \cup \mc N_{\ad^\fr} ^i$, $\mc N_2\!=\!\mc N_{\ac^\oo}^i \! \cup \mc N_{\ad^\oo}^i$, and $\mc N_3\!=\!\mc N_{\ac^\oo}^i$, or  
		\item \label{cond:cov.dominated}$\mc N_1\!=\!\mc N_{\ad}^i$, $\mc N_2\!=\!\mc N_3\!=\!\mc N_{\ac}^i $.  
	\end{enumerate}
\end{condition}
Broadly speaking, Cond.~\ref{cond:topology.both} ensures that every node that without frequency response synchronizes to at least one node with frequency response (see Sec.~\ref{sec:ac.subnetwork.topology.cond.expanation} for a detailed discussion). Moreover, by requiring Cond.~\ref{cond:topology.both} to hold after deleting a given number of arbitrary nodes or edges, robustness to  a loss of lines or generators can be established (see \cite[Assump. 1]{SG21}).
\subsection{Small signal stability}
Finally, to prove nominal stability (i.e., $P_d=\mathbbl{0}_{n_d}$) of \eqref{eq:matrix.A}, we define $x_\delta \coloneqq (\eta_\delta,\omega_\delta,v_\delta,P_{\fr,\delta}) \in \R^n$ with $n\coloneqq |\mc E_\ac|+ |\mc N_\ac| + |\mc N_\cc|+|\mc N_\dc|+|\mc N_\fr|$ and restrict \eqref{eq:matrix.A} to $P_{\zs,\delta}\!=\!\mathbbl{0}_{|{\mc N}_\zs|}$. Next, we define the LaSalle function $V\coloneqq x_\delta^\T \mc M x_\delta$ with  $\mc M \coloneqq \tfrac{1}{2 } \blkdiag \{\mc W_\ac , M, \tilde{K}_\omega C, \tilde{K} T_\fr \}$, $\tilde{K} \!\coloneqq\! (\mc I_{\fr,\ac}\mc I_{\fr,\ac}^\T +\mc I_{\fr,\dc} \mc I_{\dc} \tilde{K}_\omega  \mc I_\dc ^\T \mc I_{\fr,\dc}^\T) K_\g^{-1}$, and $\tilde{K}_\omega \!\coloneqq\! \diag \{k_{\omega}^i I_{|\mc N^i_\cc \cup \mc N^i_\dc|}\}_{i=1}^{N_\dc}$ with $k_{\omega}^i$ from Cond.~\ref{assump:identical.k.theta.gains}. First, we bound $\ddt V$ under Cond. \ref{assump:identical.k.theta.gains}-\ref{lemma:more.conservative.gains.assumption}.
\begin{proposition} \label{prop:derivative.expression} \textbf{(LaSalle function)} 
Under Cond. \ref{assump:identical.k.theta.gains}-\ref{lemma:more.conservative.gains.assumption} the function $V$ is positive definite and for $P_d=\mathbbl{0}_{n_d}$ its time derivative along the trajectories of \eqref{eq:matrix.A} restricted to $P_{\zs,\delta}=\mathbbl{0}_{|{\mc N}_\zs|}$  satisfies $\ddt V\! = \!-\!\tilde{x}_\delta^\T \mc V\tilde{x}_\delta \!-\! \tfrac{1}{2} v_\delta^\T ( \tilde{K}_{\omega} \Xi \! + \!  \Xi \tilde{K}_{\omega} )v_\delta  \!-\!   \omega_\delta^\T \mc I_\w^\T K_{\w} \mc I_\w \omega_\delta \! -\! P_{\fr,\delta}^\T \! \tilde{K} P_{\fr,\delta} \! \leq \!  0$, with $\tilde{x}_\delta \coloneqq (\mc I_{\ad} \mc B_\ac \mc W_\ac \eta_\delta,v_\delta)$, $\Xi \coloneqq \mc I_\dc^\T \mc I_\pv^\T K_\pv \mc I_\pv \mc I_\dc$, and 
\begin{align*}
\mc V \coloneqq \left [ \begin{smallmatrix} K_p  \mc I_\da C ^{-1} \mc I_\da^\T & \frac{1}{2} K_p \mc I_\da C^{-1} L_\dc \\ 
\star & \frac{1}{2}(\tilde{K}_\omega L_\dc + L_\dc \tilde{K}_\omega) \end{smallmatrix} \right]. 
\end{align*}
\end{proposition}
A proof is provided in~App.~\ref{app:proof}. To characterize the set of asymptotically stable states, the following proposition characterizes the largest invariant set $\bar{\mc S}$ (i.e.,  $x_\delta(0)\!\in\!\bar{ \mc S} \Rightarrow x_\delta(t)\!\in\!\bar{\mc S}$ for all $t\!\geq\!0$) contained in $\mc S \coloneqq \{x_\delta \in  \R^n \vert \ddt V(x_\delta(t)) =0 \}$. 
\begin{proposition} \label{prop:max.inv.set} \textbf{(Largest invariant set)}
Consider the dynamics \eqref{eq:matrix.A} with $P_d=\mathbbl{0}_{n_d}$ and restricted to $P_{\zs,\delta}\!=\! \mathbbl{0}_{|{\mc N}_\zs|}$. If Cond.~\ref{assump:identical.k.theta.gains}-\ref{assump:onestab} and Cond.~\ref{cond:topology.both} hold, then the origin is the largest invariant set contained in $\bar{\mc S}$.
\end{proposition}
A proof sketch is provided in~App.~\ref{app:proof}. Finally, we are ready to state our main stability result.
\begin{theorem} \label{thm:main.theorem}\textbf{(Stability of hybrid ac/dc power systems)} 
If Cond.~\ref{assump:identical.k.theta.gains}-\ref{assump:onestab} and Cond.~\ref{cond:topology.both} hold, then \eqref{eq:matrix.A} with $P_d=\mathbbl{0}_{n_d}$ is asymptotically stable with respect to the origin.
\end{theorem}
The theorem follows from Prop.~\ref{prop:derivative.expression}, Prop.~\ref{prop:max.inv.set}, and using the same steps as in the proof of \cite[Thm. 1]{SG21}. Linearity and asymptotic stability of \eqref{eq:matrix.A} directly implies the following corollary.
\begin{corollary}\textbf{(Stability under constant disturbances)}
	Under Cond.~\ref{assump:identical.k.theta.gains}-\ref{assump:onestab}, Cond.~\ref{cond:topology.both}, and $\ddt P_d=0$, \eqref{eq:matrix.A} is exponentially stable with respect to $x^{\sst} = -A^{-1} B P_d$.
\end{corollary}
Next, we analyze the steady state frequency for constant $P_d$.

\subsection{Steady-state analysis}\label{sec:ss.analysis}
In steady-state (i.e., $\ddt x_{\delta,l}=0$) the derivative term in the universal dual-port GFM control \eqref{eq:control.law} vanishes and the steady-state satisfies $\omega_{\delta,l}^\sst=k_{\omega,l}v_{\delta,l}^\sst$. We recall that the droop coefficient $\kappa_{p,l}\!\! \coloneqq \!\! 1/k_{\g,l}$ of SGs describes its steady-state frequency deviation $\omega^\sst_{\delta,l}$ as a function of the steady-state power injection $P^\sst_{\fr,\delta,l}$ (i.e., $\omega^\sst_{\delta,l}= -\kappa_{p,l} P^\sst_{\fr,\delta,l}$). 

Analogously,  $\kappa_{P,l}\in \R_{>0}$ denotes the \emph{effective droop coefficient} of renewable generation, i.e., $\omega^\sst_{\delta,l}= -\kappa_{P,l} P^\sst_{\fr,\delta,l}$. Using the models from Sec.~\ref{sec:generation.models}, we obtain $\kappa_{P,l}=k_{\omega,l}/k_{\pv,l}$ for curtailed PV, $\kappa_{P,l}=k_{\omega,l}/k_{\g,l}$ for a DC source with $k_{\g,l}>0$, and $\kappa_{P,l}=k_{\omega,k}/(k_{\omega,l}(k_{\g,l}+k_{\w,l}))$ for curtailed PMSG WTs~\cite{LSG22}.
Notably, the droop coefficient $\kappa_{P,l}$ is typically prescribed by, e.g., a system operator and/or grid code and can be met by adjusting $k_{\omega,l}$ and the (curtailed) operating point, while $k_{p,l}$ can be used to shape the transient response.

Next we show that, under mild assumptions, the steady-state frequency of the overall system is determined by the droop coefficients $\kappa_{P,l}$ and the generation/load mismatch $P_d$.
\begin{proposition} \label{prop:ss.sync.freq} \textbf{(Quasi-synchronous hybrid ac/dc network)} 
Assume that $g_{lk}^\dc \rightarrow \infty$ for all $(l,k)\in \mc E_\dc$ and define $D\coloneqq \sum\nolimits_{l \in \mc N_{\fr} \cup \mc N_{\pv} \cup \mc N_\w} \kappa_{P,l}^{-1}$. Then, under Cond.~\ref{assump:identical.k.theta.gains} and Cond.~\ref{cond:topology.both} it holds that $\omega_{\delta,l}^\sst = -\mathbbl{1}^T_{n_d} P_d/D$  for all $l \in \mc N_\ac \cup \mc N_\cc.$
\end{proposition}
A proof is provided in~App.~\ref{app:proof}. A key feature of  universal dual-port GFM control \eqref{eq:control.law} is that the system admits a quasi-synchronous steady-state with identical (scaled) frequency deviations if dc line losses are negligible (i.e., $g_{lk}^\dc \rightarrow \infty$). Therefore, the signals that indicate power imbalances (i.e., $\omega_{\delta,l}$ and $v_{\delta,l}$) synchronize up to scaling by $k_{\omega,l}$ and the overall system response can easily be predicted based on the droop coefficients $\kappa_{P,l}$ of each power source. This feature is also be beneficial for system planing purposes. Moreover, this result highlights the ability of universal dual-port GFM control \eqref{eq:control.law} to induce a system-wide primary frequency response that rebalances short-term energy storage elements (e.g., SG and WT kinetic energy) and their corresponding frequencies/dc voltages after a load disturbance. Despite the lack of VSC power setpoints, the nominal operating point of the overall system is fully determined through the setpoints for VSC frequency and dc voltage and setpoints of the power generation.

In contrast, using power-balancing dual-port GFM control \eqref{eq:control.law.pbdp} the steady-state frequencies of different ac subnetworks exhibit a complicated dependence on the power setpoints and droop coefficients of both the VSCs and power sources and do not admit a quasi-synchronous steady state. This can result in counter-intuitive responses when post-contingency power flows significantly differ from the VSC power set-point~\cite{GSP+21}. 
\section{Interpretation of stability conditions and control tuning} \label{sec:tuning}
In this section, we we discuss the consequences of the stability conditions on the VSC control gains for (i) control tuning to meet steady-state and transient specifications, and (ii) selecting operating points of renewable generators. Next, we illustrate our stability conditions in the context of the application examples presented in Sec.~\ref{sec:illustrative example}. Finally, we discuss the interpretation of the conditions on the ac subnetwork topologies and their relationship to results in the literature.

Before proceeding, we stress that, while our stability conditions are largely independent of the network parameters and operating points of renewables, the frequency regulation performance is not. In other words, parameter variations can greatly impact system performance and more detailed comparisons with well-known controls are needed to evaluate these aspects (see, e.g., \cite{LG23} for results on PMSG WTs)

\subsection{Steady-state response and Condition~\ref{assump:identical.k.theta.gains}}\label{sec:cond1ss}
As discussed in Sec.~\ref{sec:ss.analysis}, the steady-state response of the system is determined by the VSC control gain $k_{\omega,l}$ and sensitivities of the power sources. Thus, $k_{\omega,l}$ should be tuned to satisfy stability and steady-state specifications. 

For each dc subnetwork, $k_{\omega,l}$ can always be chosen such that Cond.~\ref{assump:identical.k.theta.gains} holds and the frequencies imposed on neighboring ac subnetworks are consistent in steady-state. To understand why Cond.~\ref{assump:identical.k.theta.gains} is often necessary, consider bus~30 and bus~60 of the hybrid ac/dc system shown in Fig.~\ref{fig:grid} at a hypothetical steady-state with $v_{\delta,30} = v_{\delta,60} \neq 0$. Then $k_{\omega,30} \neq k_{\omega,60}$, then $\omega_{\delta,30} \neq \omega_{\delta,60}$ despite DC~2 being in steady-state, i.e., AC~2 cannot be at a synchronous steady state. This results in circulating power flows between ac subnetworks that are interfaced through multiple dc subnetworks (see \cite{SG2023}, \cite[Ex.~2]{SG21}, and \cite[Sec.~V-D]{GSP+21} for details). However, Cond.~\ref{assump:identical.k.theta.gains} can be lifted for ac subnetworks that are only connected through one dc subnetwork, e.g., PMSG WTs (cf. \cite[Thm.~1]{LSG22}) or two ac systems connected by a point-to-point HVDC link. 

Moreover, for curtailed renewable generation and dc power sources with $k_{\g,l}>0$, the effective droop gain $\kappa_{P,l}$ crucially hinges on $k_{\omega,l}$ (see Sec.~\ref{sec:ss.analysis}). Thus,  $k_{\omega,l}$ needs to be tuned to meet steady-state droop specifications for each device (e.g., provided by the system operator  or determined through market mechanisms). In particular, the effective droop gain $\kappa_{P,l}$ of renewables depends on $k_{\omega,l}$ and the sensitivities (i.e., $k_{\pv,l}$, $k_{\w,l}$, $k_{\beta,l}$ defined Sec.~\ref{sec:generation.models}) that in turn depend on the operating conditions (i.e., wind speed) and curtailment (see Sec.~\ref{sec:op.point}). In other words, $k_{\omega,l}$ may have to be adjusted as the operating conditions or curtailment changes. For example, as discussed in \cite{LSG22}, Cond.~\ref{assump:identical.k.theta.gains} can be relaxed for PMSG WTs and the ratio $k_{\omega,l}/k_{\omega,k}$ can be used to match a given $\kappa_{P,l}$.

In contrast, if i) renewables are operated at the MPP, or ii) VSC is an interlinking converter connecting ac and dc networks, or iii) VSC is a part of the HVDC link, we obtain the upper bound $k_{\omega,l}\leq \Delta\omega_{l}^\text{max}/\Delta v_{l}^\text{max}\eqqcolon k_{\omega,l}^{\max}$. The upper $k_{\omega,l}^{\max}$ is based on the largest expected frequency deviation (e.g., the boundary of the nominal operating range) $\Delta\omega_{l}^\text{max}$ and the largest acceptable dc voltage deviation $\Delta v_{l}^\text{max}$. In case of a PMSG WT, the steady-state relationships between the VSC frequencies, dc voltages, and physical limitations of the WT have to be considered when selecting $k_{\omega,l}$ (for details see~\cite{LSG22}).

\subsection{Transient response and Condition~\ref{lemma:more.conservative.gains.assumption}}\label{sec:cond2transient}
Next, given $k^i_\omega$ determined by steady-state specifications and Cond.~\ref{assump:identical.k.theta.gains}, it remains to select $k_{p,l}$ to satisfy Cond.~\ref{lemma:more.conservative.gains.assumption} to ensure stability of the closed-loop system.

In particular, within each dc subnetwork Cond.~\ref{lemma:more.conservative.gains.assumption} quantifies the set of stabilizing control gains that crucially hinges on the effective (parallel) dc network resistance $r^\dc_{\text{eq},l}$ relative to node $l \in \mc N_{\da}^i$ and (scaled) dc-link capacitance $c_l=C_l v^\star_l$. In particular, a larger dc capacitance $C_l$ or increased dc voltage setpoint $v^\star_l$ imply more energy stored in the dc capacitor and larger $c_l$. Thus, Cond.~\ref{lemma:more.conservative.gains.assumption} shows that increased energy stored in the dc link, increased damping in the dc network (i.e., larger effective resistance $r^\dc_{\text{eq},l}$), and larger steady-state gain $k_{\omega,l}$  allows for faster voltage phase angle control (larger control gain $k_{p,l}$) and increased oscillation damping. In contrast, decreasing dc line losses or dc bus capacitance both require decreasing $k_{p,l}$, i.e., $k_{p,l}\!\to\! 0$ as $r^\dc_{l,k}\! \to \! 0$ or $c_l \!\to \! 0$. 

\subsection{Conditions for HVDC, LFAC, and renewables}
Next, we discuss how Cond.~\ref{lemma:more.conservative.gains.assumption} can be relaxed and interpreted for common network topologies and converter interfaced renewable generation.
 \subsubsection{HVDC} If HVDC line losses are negligible, the frequencies of HVDC VSCs interconnecting ac networks satisfy $\omega_{\delta,l}^\sst/\omega_{\delta,k}^\sst\!=\! k_{\omega,l}/k_{\omega,k}$. In other words, Cond.~\ref{assump:identical.k.theta.gains} is necessary if the ac areas are synchronous. For point-to-point HVDC, Cond.~\ref{assump:identical.k.theta.gains} can be relaxed, allowing for different $k_{\omega,l}$ gains, if the ac terminals of the HVDC link are connected to asynchronous ac areas. Additionally, for point-to-point HVDC, Cond.~\ref{lemma:more.conservative.gains.assumption} simplifies to $k_{p,l}/k_{\omega,l}  \!<\! 2 c_l r_{lk}^\dc$, i.e., decreased losses and dc-link capacitance require smaller $k_{p,l}$ because $k_{\omega,l}$ is typically fixed by steady-state specifications. Moreover, the nominal power across HVDC can be scheduled through $v_l^\star$ and $v_k^\star$ \cite{GSP+21}. 
 \subsubsection{LFAC} LFAC systems are modeled by combining VSCs with back-to-back dc connection. For the LFAC system in Fig.~\ref{fig:sources}~f) connecting two asynchronous ac systems, Cond.~\ref{assump:identical.k.theta.gains} is not required and the mapping of steady-state frequency deviations across the LFAC connection (with VSCs numbered sequentially) is given by $\omega_{\delta,l}^\sst/\omega_{\delta,l+3}^\sst=(k_{\omega,l}k_{\omega,l+2})/(k_{\omega,l+1}k_{\omega,l+3})$. In this case, using the same steps as in \cite[Sec.~V]{LSG22}, it can be shown that Cond.~\ref{lemma:more.conservative.gains.assumption} can be replaced by $k_{p,l+k}>0$, $k\in \N_{[0,3]}$ for each LFAC converter.  
 \subsubsection{DAB} Dual active bridge DC/DC converters are modeled by two VSCs in back-to-back ac connection. In steady state, the $l$th and $k$th VSC dc voltages (i.e., terminals of the DAB) satisfy $v_{\delta,l}^\sst/v_{\delta,k}^\sst=k_{\omega,k}/k_{\omega,l}$ and, in our framework, Cond.~\ref{lemma:more.conservative.gains.assumption} can, in general, not be relaxed.
 \subsubsection{Converter interfaced renewable generation}
 When operating converter-interfaced renewable generation at a curtailed operating point, \eqref{eq:control.law} provides an inertia response and grid-forming primary frequency control with effective droop coefficients (see Sec.~\ref{sec:ss.analysis}) similar to ac-GFM/dc-GFL control. In contrast, if renewables are operated at their MPP, their sensitivity to frequency and dc voltage deviations is negligible (see Fig.~\ref{fig:operating_points}). Hence, the proposed control approximately tracks the MPP like ac-GFL/dc-GFM control. In the following, we examine bounds on the derivative gain of renewable generation depending on the operating point (i.e., curtailed vs. MPP).
 
 Regardless of the operating point, Cond.~\ref{lemma:more.conservative.gains.assumption} can be replaced by $k_{p,l}\!>\!0$ for \emph{PMSG WTs} \cite{LSG22}. Controllable dc sources (e.g., battery energy storage) and PV can often be modeled as directly feeding into the VSC dc link, i.e., $P_{\delta,\dc,l}=-P_{\delta,l}$, and Cond.~\ref{lemma:more.conservative.gains.assumption} simplifies. For dc sources with $k_{\g,l}=0$ such as PV operating at its MPP, Cond.~\ref{lemma:more.conservative.gains.assumption} simplifies to $k_{p,l}\!>\!0$. In contrast, for dc sources with $k_{\g,l}>0$ such as curtailed PV, Cond.~\ref{lemma:more.conservative.gains.assumption} simplifies to $k_{p,l}\!<\!4 c_l \kappa_{p,l}$.

\subsection{Conditions on ac subnetwork topologies}\label{sec:ac.subnetwork.topology.cond.expanation}
Cond.~\ref{assump:onestab} ensures that at least one device (including conventional generators) can respond to disturbances in the grid. Without this fundamental assumption, the power system cannot be stable. Thus, while the capabilities of renewables and availability of reserves change over time, at least one device must respond to disturbances in the grid at any given time.

Next, Cond.~\ref{cond:topology.both}~\textit{\ref{cond:few.machines.zs}} requires that every ac subnetwork contains connections through which devices $\mc N_{\ac^\oo}^i \! \cup \mc N_{\ad^\oo}^i$ without frequency response  synchronize to devices $\mc N_{\ac^\fr}^i\! \cup \mc N_{\ad^\fr}^i$ with frequency response. This implicitly requires sufficiently many devices with frequency response (i.e., $|\mc N_{\ac^\fr}^i\! \cup \mc N_{\ad^\fr}^i| \geq |\mc N_{\ac^\oo}^i \! \cup \mc N_{\ad^\oo}^i|$). Alternatively, Cond.~\ref{cond:topology.both}~\textit{\ref{cond:cov.dominated}} requires connections through which machines $\mc N_{\ac}^i$ synchronize to VSCs $\mc N_{\ad}^i$, i.e., implicitly requires sufficiently many VSCs (i.e., $\mc N_{\ad}^i \geq \mc N_{\ac}^i$). Notably, Cond.~\ref{cond:topology.both} only requires partial knowledge of the ac subnetwork topology and does not require knowledge of line susceptances. While Cond.~\ref{cond:topology.both}~\textit{\ref{cond:few.machines.zs}} implicitly depends on the operating point and curtailment of renewables (i.e., $k_{\g,l}$), Cond.~\ref{cond:topology.both}~\textit{\ref{cond:cov.dominated}} is independent of the operating point of renewables. In other words, if an ac subnetwork is converter-dominated (i.e., $\mc N_{\ad}^i \gg \mc N_{\ac}^i$), Cond.~\ref{cond:topology.both}~\textit{\ref{cond:cov.dominated}} can cover a wide range of operating points.

Condition~\ref{cond:topology.both} contains well-known results in the literature for multi-machine and/or multi-converter systems as special cases. Notably, Cond.~\ref{cond:topology.both}~\textit{\ref{cond:few.machines.zs}} and Cond.~\ref{cond:topology.both}~\textit{\ref{cond:cov.dominated}} are not mutually exclusive. Generally, Cond.~\ref{cond:topology.both}~\textit{\ref{cond:few.machines.zs}} is directly applicable to ac subnetworks with \emph{few SMs without frequency control} and Cond.~\ref{cond:topology.both}~\textit{\ref{cond:cov.dominated}} is directly applicable to \emph{converter dominated} ac subnetworks. The following corollary presents cases for which stability can be guaranteed without any conditions on the system topology. 
\begin{corollary}\textbf{(Simplified stability condition)}  \label{corl:ac.cond.simple} 
	For each ac subnetwork, Cond.~\ref{cond:topology.both} holds if the ac subnetwork 
	\begin{enumerate}[label=\roman*)]
	\item \label{corl.ac.leg.mach} only contains SGs (i.e., $k_{\g,l}>0$), or
	\item \label{corl.ac.only.conv} only contains VSCs, or	
	\item \label{corl.ac.mixed.conv.legacy} only contains  VSCs and SGs (i.e., $k_{\g,l}>0$), or
	\item \label{corl.ac.mixed.conv.PMSG.WT} only contains  VSCs and one SM (e.g., PMSG WT), or 
	\item \label{corl.ac.mix.conv.SC} only contains VSCs and one SC.
\end{enumerate}
\end{corollary}
In particular, Cor.~\ref{corl:ac.cond.simple}~\textit{\ref{corl.ac.leg.mach}} recovers well-known results for multi-machine systems and immediately follows from Cond.~\ref{cond:topology.both}~\textit{\ref{cond:few.machines.zs}} with $\mc N_2=\mc N_3=\emptyset$. In contrast, Cor.~\ref{corl:ac.cond.simple}~\textit{\ref{corl.ac.only.conv}} follows from Cond.~\ref{cond:topology.both}~\textit{\ref{cond:cov.dominated}} with $\mc N_2=\mc N_3=\emptyset$. Notably, topology-independent stability results for networks of VSCs with controllable dc source (e.g., \cite{MDPS+17}) are a special case of Cor.~\ref{corl:ac.cond.simple}~\textit{\ref{corl.ac.only.conv}}. Corollary~\ref{corl:ac.cond.simple}~\textit{\ref{corl.ac.mixed.conv.legacy}} follows from Cond.~\ref{cond:topology.both}~\textit{\ref{cond:few.machines.zs}} with $\mc N_3=\emptyset$ and shows that a mix of SGs and VSCs is stable independently of the topology.

For ac subnetworks as in Cor.~\ref{corl:ac.cond.simple}~\textit{\ref{corl.ac.mixed.conv.PMSG.WT}} and Cor.~\ref{corl:ac.cond.simple}~\textit{\ref{corl.ac.mix.conv.SC}}, Cond.~\ref{cond:topology.both}~\textit{\ref{cond:cov.dominated}} is trivially satisfied since $\mc N_2$ in Cond.~\ref{cond:topology.both}~\textit{\ref{cond:cov.dominated}} only contains one node that is connected to at least one node in $\mc N_1$.  However, if the ac subnetwork contains more than one SC, Cond.~\ref{cond:topology.both} has to be checked. Notably, much like ac-GFL control, SCs ultimately need to synchronize with a device with stable frequency dynamics. We now apply Cor.~\ref{corl:ac.cond.simple} to illustrate our conditions for the hybrid dc/ac system shown in Fig.~\ref{fig:grid}. 

\begin{example}\textbf{(Stability conditions for hybrid dc/ac system)} 
	Because multiple SGs provide primary frequency control (i.e., $k_{\g,l}>0$), Cond.~\ref{assump:onestab} holds for the system shown in Fig.~\ref{fig:grid}. Moreover, by inspection Fig.~\ref{fig:grid}  and the ac subnetwork graphs in Fig.~\ref{fig:subnetworks}~a), Cor.~\ref{corl:ac.cond.simple}~\textit{\ref{corl.ac.mixed.conv.legacy}} holds for AC~1 and AC~3, Cor.~\ref{corl:ac.cond.simple}~\textit{\ref{corl.ac.mix.conv.SC}} holds for AC~2, and Cor.~\ref{corl:ac.cond.simple}~\textit{\ref{corl.ac.mixed.conv.PMSG.WT}} holds for AC~4. Thus, the system in Fig.~\ref{fig:grid} is stable if the VSC control gains satisfy Cond.~\ref{lemma:more.conservative.gains.assumption}.
	\end{example}
\section{Case study: Switched Simulation} \label{sec:case.study}
\begin{figure}[t!]
	\centering
	\includegraphics[trim=2mm 0 5mm 0, clip,width=0.75\columnwidth]{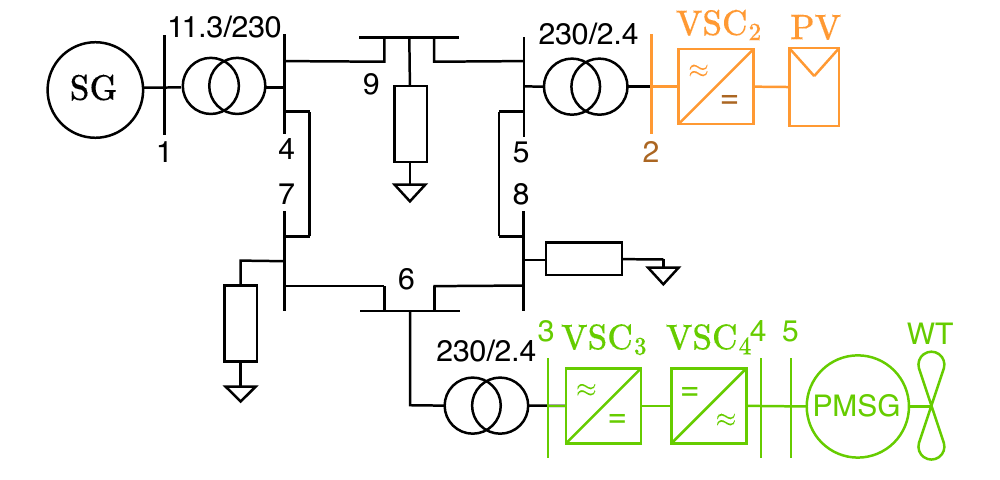} 
	\caption{System containing a PMSG WT, single-stage PV system, and a SG.\label{fig:small.case.study}} 
\end{figure}

We first use the system depicted in Fig.~\ref{fig:small.case.study} to illustrate dual-port GFM control applied to two-level VSCs with IGBT switch model and $10$~kHz PWM in an electromagnetic transient (EMT) simulation using Simscape Electrical in MATLAB Simulink. Our simulations use a discrete-time implementation of the derivative-free implementation \eqref{eq:control.pi.interpretation} with sampling rate of $10$~kHz and computation delay of $1e^{-4}$~s. The system base power and frequency are $S_b=100$~MW and $f_b=50$~Hz and the base values for the different voltage levels can be found in~\cite[Table I]{SG21}. The SG is modeled using an $8^\text{th}$ order model with standard exciter, automatic voltage regulator, multiband power system stabilizer, and a first-order turbine/governor model with $5\%$ speed droop (see~\cite[Table I]{TGA+20}). 

The WT represents an aggregate of ten identical $5$~MW PMSG WTs (see~\cite{LSG22} for details) modeled as a single PMSG WT with detailed control implementation (i.e., cascaded inner loops, PWM modulation, pitch servo model and control, etc.). Similarly, a single-stage PV system is modeled by aggregating $5000$ parallel strings of $90$ AUO PM060MBR modules and interfacing the aggregate PV module model through a two-level VSC with detailed control implementation (e.g., cascaded inner loops, PWM modulation). The VSC parameters and aggregation procedure can be found in \cite{TGA+20}. Moreover, for brevity, plant level controls are not implemented. Additionally, we use dynamical models of the low-voltage/high-voltage, medium-voltage/high-voltage~\cite[Table I]{TGA+20} and high-voltage/high-voltage~\cite{SG21} transformers. The ac lines and dc cables are modeled using the standard $\pi$-line dynamics with parameters as in~\cite{SG21}. The voltage magnitude of the VSCs is controlled through standard GFM $Q\!-\!V$ droop of $1\%$~\cite{DSF2015} for the PV system and controlled to zero using PI $Q-V$ droop for the WT. A notch filter at the PWM frequency as well as lowpass filters with $2$~kHz cutoff frequency are used to remove switching ripple from the dc voltage measurements used in \eqref{eq:control.pi.interpretation}. For $l \in \{2,3,4\}$, the proportional and integral gains of the voltage PI control are $k_{p,l}^{\text{v}}=0.79$~p.u. and $k_{i,l}^{\text{v}}=0.69$~p.u. The integral gain of the current PI control is $k_{i,l}^\text{i}=0.79$~p.u. for $l \in \{2,3,4\}$. The proportional gain of the current PI control is $k_{p,l}^\text{i}=0.9$~p.u. for $l\in \{3,4\}$ and $k_{p,l}^\text{i}=1.3$~p.u. for $l=2$. The MPP of the PV system corresponds to $1.37$~p.u. under standard conditions.

The remainder of this section presents simulation results for three scenarios that demonstrate grid support by curtailed renewables and approximate MPP.

\subsection{Curtailed PV \& WT}\label{sec:pv.wt.curtailed}
\begin{figure}[h]
	\centering
	\includegraphics[width=1\columnwidth]{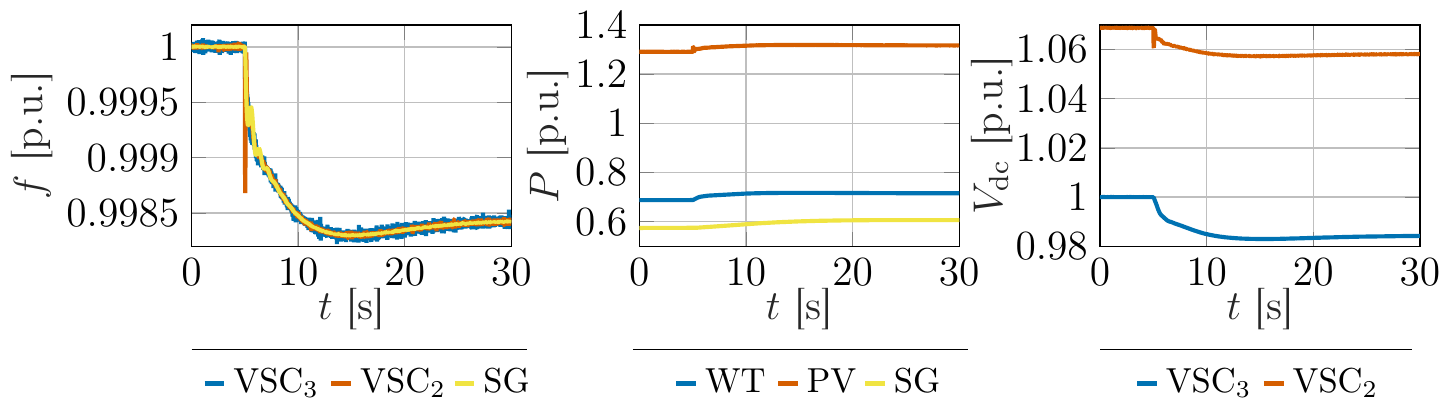} 
	\caption{Frequency of the SG, VSC$_2$ and VSC$_3$, power generated by SG, and curtailed PV and WT, and dc voltage of the VSC$_2$ and VSC$_3$ during disturbance of $0.075$~p.u. at $t=5$~s.\label{fig:pv.wt.curtailed}}
\end{figure}

To show that dual-port GFM control can provide grid support, we first consider curtailed WT and PV (see Sec.~\ref{sec:op.point}). The simulation starts from the nominal operating point $P_{\text{SG}}^\star=0.575$~p.u., $P_{\text{PV}}^\star=95\% P_{\text{PV}}^\text{MPP}$ and $P_{\text{WT}}^\star=91.73\% P_{\text{WT}}^\text{MPP}$ with $\kappa_{\text{SG}}=\kappa_\text{PV}=5\%$ and $\kappa_\text{WT}=3.33\%$. A load step of $0.075$~p.u. occurs at $t=5$~s. Figure~\ref{fig:pv.wt.curtailed} shows the frequency of the SG, VSC$_2$ (PV), and VSC$_3$ (WT grid side converter), as well as the power generation of the WT and PV module and the dc voltage of VSC$_2$ and VSC$_3$. The WT and PV system respond to the load step according to their effective droop gains (see Sec~\ref{sec:ss.analysis}) illustrating that dual-port GFM control provides grid support when renewable generation operates at a suitable curtailed operating point.

\begin{figure}[h]
		\centering
		\includegraphics[width=1\columnwidth]{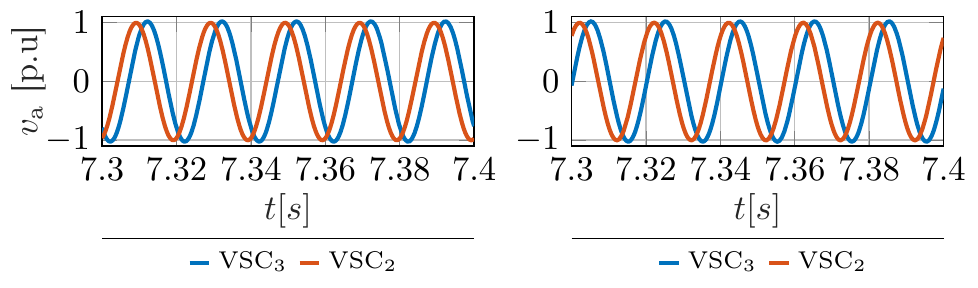}  
		\caption{Phase voltage $v_\text{a}$ at the VSC filter capacitor for the scenarios in Sec.~\ref{sec:pv.wt.curtailed} (left) and Sec.~\ref{sec:pv@mpp} (right). \label{fig:pv.wt.curtailed_vas}	}
\end{figure}

Moreover, as illustrated in Fig.~\ref{fig:pv.wt.curtailed}, the dc voltage is stabilized near the nominal operating point. While it can be seen that the internal frequency of the dual-port control is impacted by the dc voltage ripple, the voltage at the converter filter capacitor does not exhibit any distortion (see Fig.~\ref{fig:pv.wt.curtailed_vas}). This highlights that the dc voltage switching ripple does not deteriorate the ac voltage waveform generated by \eqref{eq:control.law}.

\subsection{PV at the MPP and curtailed WT}\label{sec:pv@mpp}
\begin{figure}[h]
	\centering
	\includegraphics[width=1.01\columnwidth]{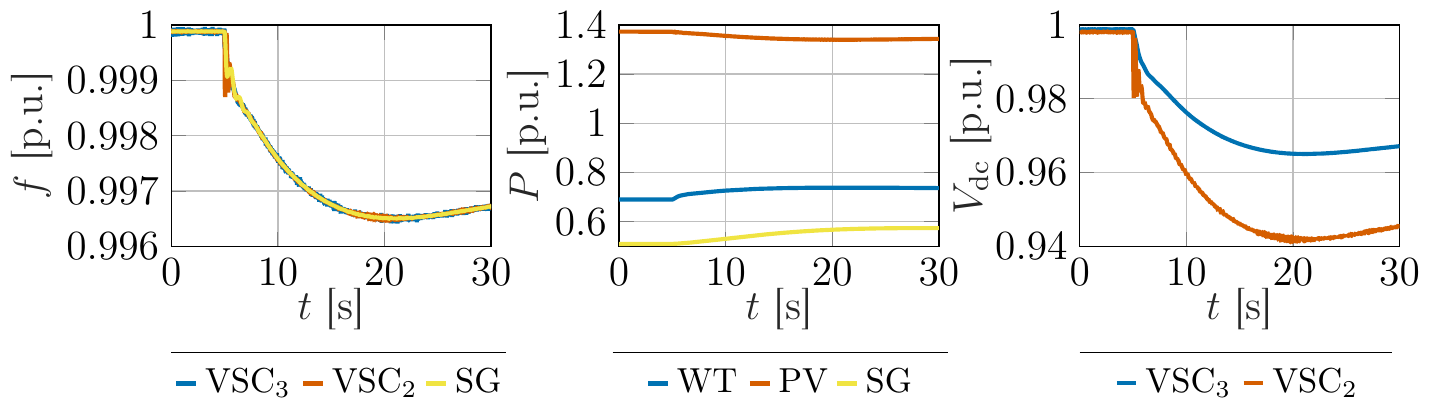} 
	\caption{Frequency of the SG, VSC$_2$ and VSC$_3$, power generated by SG, PV at MPP and curtailed WT, and dc voltage of the VSC$_2$ and VSC$_3$ during disturbance of $0.075$~p.u. at $t=5$~s. \label{fig:pv.MPP}}
\end{figure}

Next, we illustrate that dual-port GFM control can approximately track the MPP. To this end, we change the operating point of the solar PV converter to the MPP. In this case the simulation starts from the nominal operating point $P_{\text{SG}}^\star=0.5075$~p.u., $P_{\text{PV}}^\star=P_{\text{PV}}^\text{MPP}$ and $P_{\text{WT}}^\star=91.73\% P_{\text{WT}}^\text{MPP}$ with $\kappa_{\text{SG}}=5\%$ and $\kappa_\text{WT}=3.33\%$. In contrast, $k_{\omega,2}=0.2$~p.u. and $\kappa_\text{PV}=/$. A load step of $0.075$~p.u. occurs at $t=5$~s. Figure~\ref{fig:pv.MPP} illustrates the frequency of the SG and VSCs, the power output of the PV and WT, and VSC dc voltages. As in Sec.~\ref{sec:pv.wt.curtailed}, the WT responds to the load step according to its effective droop gain. However, the PV system remains near the MPP. While the dc voltage of the PV converter drops slightly, the flat PV power characteristic at the MPP only results in a negligible drop of power generation of 2.13\% below the MPP. We note that operation below the MPP voltage can be prevented by including a dc voltage limiter~\cite[Fig.~10]{CLJ19}. A detailed analysis of the dc voltage limiter and its impact on the system is seen as interesting topic for future work.

Phase a of the voltages at the filter terminal are shown in Fig.~\ref{fig:pv.wt.curtailed_vas} to verify that the switching ripple on the dc voltage does not propagate to the VSC ac voltage.
\subsection{WT at MPP and curtailed PV}\label{sec:wt@mpp}
\begin{figure}[h]
	\centering
	\includegraphics[width=1.01\columnwidth]{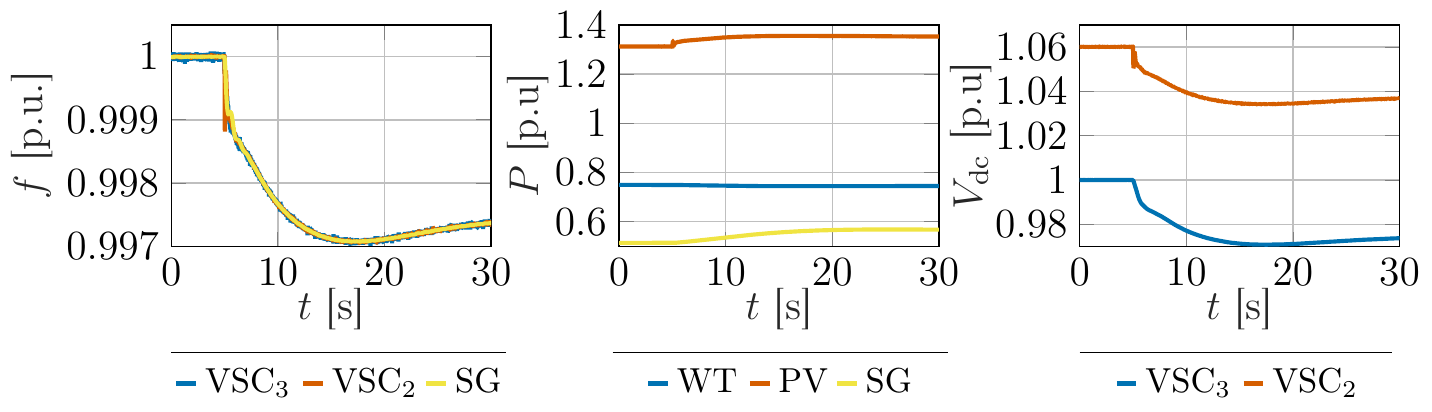} 
	\caption{Frequency of the SG, VSC$_2$ and VSC$_3$, power generated by SG, WT at MPP and curtailed PV, and dc voltage of the VSC$_2$ and VSC$_3$ during disturbance of $0.075$~p.u. at $t=5$~s.\label{fig:wt.MPP}}	
\end{figure}

\begin{figure}[h]
	\centering
	\includegraphics[width=0.99\columnwidth]{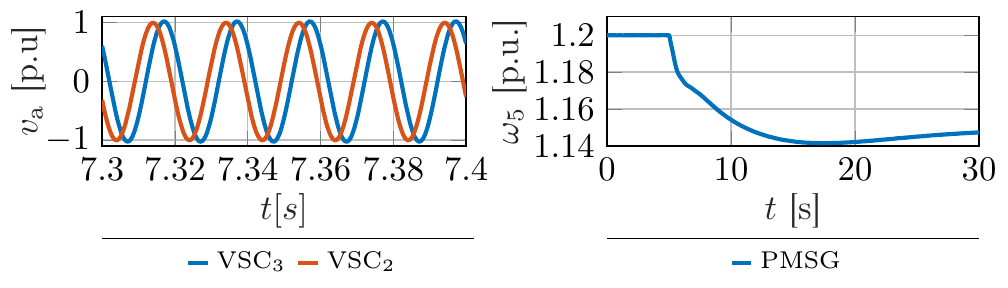}  
	\caption{Phase voltage $v_\text{a}$ at the filter capacitor of VSC$_3$ and VSC$_2$ (left) and $\omega_5$ of the PMSG (right). \label{fig:wt.mpp_vas}	}
\end{figure}

Finally, we simulate a scenario in which the WT operates at the MPP and the PV system is curtailed. The simulation starts from the nominal operating point $P_{\text{SG}}^\star=0.514$~p.u., $P_{\text{PV}}^\star=95.61\%P_{\text{PV}}^\text{MPP}$ and $P_{\text{WT}}^\star=P_{\text{WT}}^\text{MPP}$ with parameters $\kappa_{\text{SG}}=5\%$ and $\kappa_\text{PV}=5\%$. In contrast, $k_{\omega,3}=0.1$~p.u. and $k_{\omega,4}=2$~p.u and $\kappa_\text{WT}=/$. A load step of $0.075$~p.u. occurs at $t=5$~s. Figure~\ref{fig:wt.MPP} shows the frequency of the SG and VSCs, power output of the PV and WT, and VSC dc voltages. In addition, the WT rotor speed is shown in Fig.~\ref{fig:wt.mpp_vas}. The PV system responds to the load step according to its effective droop gains while the WT remains close to the MPP. The WT rotor speed drops slightly (see Fig.~\ref{fig:wt.mpp_vas}) resulting in a WT power  reduction of $0.63\%$. Avoiding operation below the MPP rotor speed can by including a rotor speed limiter~\cite[Fig.~4]{LSG22} is seen as interesting topic for future work.

We again show the phase voltage at the filter terminal in Fig.~\ref{fig:wt.mpp_vas} to verify that the switching ripple on the dc voltage does not propagate to the ac voltage.

\section{Case study: Large-scale System}

To illustrate behavior of the controller in a complex system containing renewable generation, SGs, and HVDC links we use an EMT simulation of the system in Fig.~\ref{fig:grid}. Due to the complexity and scale of the system, we use an averaged model of the two-level VSCs. The SG and SC models retain the same level of detail as in the previous section. In addition, an electrical storage system (ESS) is modeled as a controllable dc voltage source (see~\cite{TGA+20} for parameters). The parameters of the HVDC VSCs (with RLC filters) and HVDC cables can be found in~\cite{SG21}. The remaining system parameters, renewable generation MPPs, their nominal operating points, control gains, effective droop coefficients, and gains of current and voltage PI loops of the system in Fig.~\ref{fig:grid} are given in Tab.~\ref{tab:param}.

\begin{table}[b!]
	\caption{mpp, nominal operating points  \& control gains \label{tab:param}}
	\centering
	\resizebox{1\columnwidth}{!}{
		\setlength\tabcolsep{1pt}
		\bgroup
		\def\arraystretch{1.25}%
		\begin{tabular}{ c c c c c c}
			\toprule 
			\multicolumn{6}{c}{\textbf{PV$_3$} \ ({1200} parallel strings of {90} modules) \ [p.u.]} \\
			$P_3^\mpp\approx0.33$,& $P_3^\star=P_3^\mpp$, & $k_{\g,3}=0$,  &$k_{p,3}=0.001$,& $k_{\omega,3}=0.2$,& $\kappa_{P,3}= / $ \\ \midrule \midrule
			\multicolumn{6}{c}{\textbf{PV$_{12}$} ({5000} parallel strings of {90} modules) \ [p.u.]} \\
			$P_{12}^\mpp\approx1.37$, & $P_{12}^\star=94.4 \% P_{12}^\mpp$, & $k_{\g,12}\approx2.5$, &$k_{p,12}=0.001$,& $k_{\omega,12}\approx 0.12,$ & $\kappa_{P,12}= 5 \% $\\ \midrule \midrule
			\multicolumn{6}{c}{\textbf{PV$_{13}$ }({3000} parallel strings of {100} modules) [p.u.]} \\
			$P_{13}^\mpp\approx 0.92$, & $P_{13}^\star= P_{13}^\mpp$, & $k_{\g,13}=0$, &$k_{p,12}=0.001$, & $k_{\omega,13}=0.05$, & $\kappa_{P,13}= / $\\  \midrule \midrule
			\multicolumn{6}{c}{\textbf{PV$_{22}$} ({5500} parallel strings of {95} modules) \ [p.u.]} \\
			$P_{22}^\mpp\approx1.6$, & $P_{22}^\star= P_{22}^\mpp$, & $k_{\g,22}=0$, &$k_{p,22}=0.001$, & $k_{\omega,22}=0.2$, & $\kappa_{P,22}= / $\\  \midrule  \midrule
			\multicolumn{6}{c}{\textbf{PV$_{23}$} ({5700} parallel strings of {100} modules) [p.u.]} \\
			$P_{23}^\mpp\approx1.6$, & $P_{23}^\star= P_{23}^\mpp$, & $k_{\g,23}=0$, &$k_{p,23}=0.001$, & $k_{\omega,23}=0.2$, & $\kappa_{P,23}= / $\\  \midrule \midrule
			\multicolumn{6}{c}{\textbf{WT$_{71}$} ($v_\w=12$~m/s, $\beta^\star=0$) \& \textbf{DC-B2B} [p.u.]} \\
			$P_{71}^\mpp = 0.75$, & $P_{71}^\star= 90\% P_{71}^\mpp$, & $k_{\g,71}=0.6$, &\begin{tabular}{l}
				$k_{p,70}=0.015$,\\
				$k_{p,11}=0.015$,
			\end{tabular}& \begin{tabular}{l}
				$k_{\omega,70}=5$,\\
				$k_{\omega,11}=0.1$, 
			\end{tabular} &$\kappa_{p,11}= 3.33 \% $\\  \midrule \midrule
					   \multicolumn{6}{c}{\textbf{ESS$_{2}$ } (rated power 1.4~p.u.) [p.u.]} \\
		\multicolumn{2}{c}{$P_2^\star=1.1$,} & $k_{\g,2}=4$, &$k_{p,2}=0.001$,& $k_{\omega,2}\approx0.026$, & $\kappa_{P,2}= 5\% $\\  \midrule \midrule
		 \multicolumn{6}{c}{\textbf{SG$_{1}$} (rated power 1.5~p.u.), \textbf{SG$_{31}$}, \textbf{SG$_{32}$}, \textbf{SG$_{33}$} (rated power 1~p.u.) [p.u.]} \\
		 \multicolumn{4}{c}{\setlength\tabcolsep{5pt} \begin{tabular}{c c c c}$P_{1}^\star \approx 0.82$, & $P_{31}^\star\approx 0.77$, & $P_{32}^\star\approx 0.57$, & $P_{33}^\star\approx 0.37$,\end{tabular}} & $k_{\g,l}=20$, & $\kappa_{P,l}=5 \% $ \\  \midrule \midrule
		 \multicolumn{6}{c}{HVDC links: \textbf{ DC~1 }(310~km), \textbf{DC~2 }(510~km)}\\
		\multicolumn{6}{c}{$k_{p,l}=0.001\qquad k_{\omega,l}=0.2$} \\ \midrule \midrule
			  \multicolumn{6}{c}{HVDC link \textbf{DC~3} (1000~km) [p.u.]} \\
			 \multicolumn{6}{c}{$k_{p,40}=k_{p,50}=0.001 \qquad k_{\omega,40}=0.2 \qquad k_{\omega,50}=0.35$} \\ \midrule \midrule
	  \multicolumn{6}{c}{\textbf{VSC$_{l}$}, $l\in \{2,3,11,12,13,22,23,19,20,30,60,40,50 \}$ [p.u.]} \\
		  \multicolumn{6}{c}{\setlength\tabcolsep{8pt} \begin{tabular} {c c} Current PI: $k_{p,l}^\text{i}=2.1, \ k_{i,l}^\text{i}=0.79$,S & Voltage PI: $k_{p,l}^\text{v}=0.15, \ k_{i,l}^\text{v}=0.69$ \end{tabular}}  \\ \bottomrule
	\end{tabular}
\egroup	}
\end{table}

\subsection{Sequence of events}
To verify the behavior of universal dual-port GFM control, we simulate the events in Tab.~\ref{table:event.seq}. The dispatch and control gains are given in Tab.~\ref{tab:param} and Tab.~\ref{table:redisp.st}.

\begin{figure*}[t!!!]
		\centering
		\includegraphics[width=0.99\textwidth]{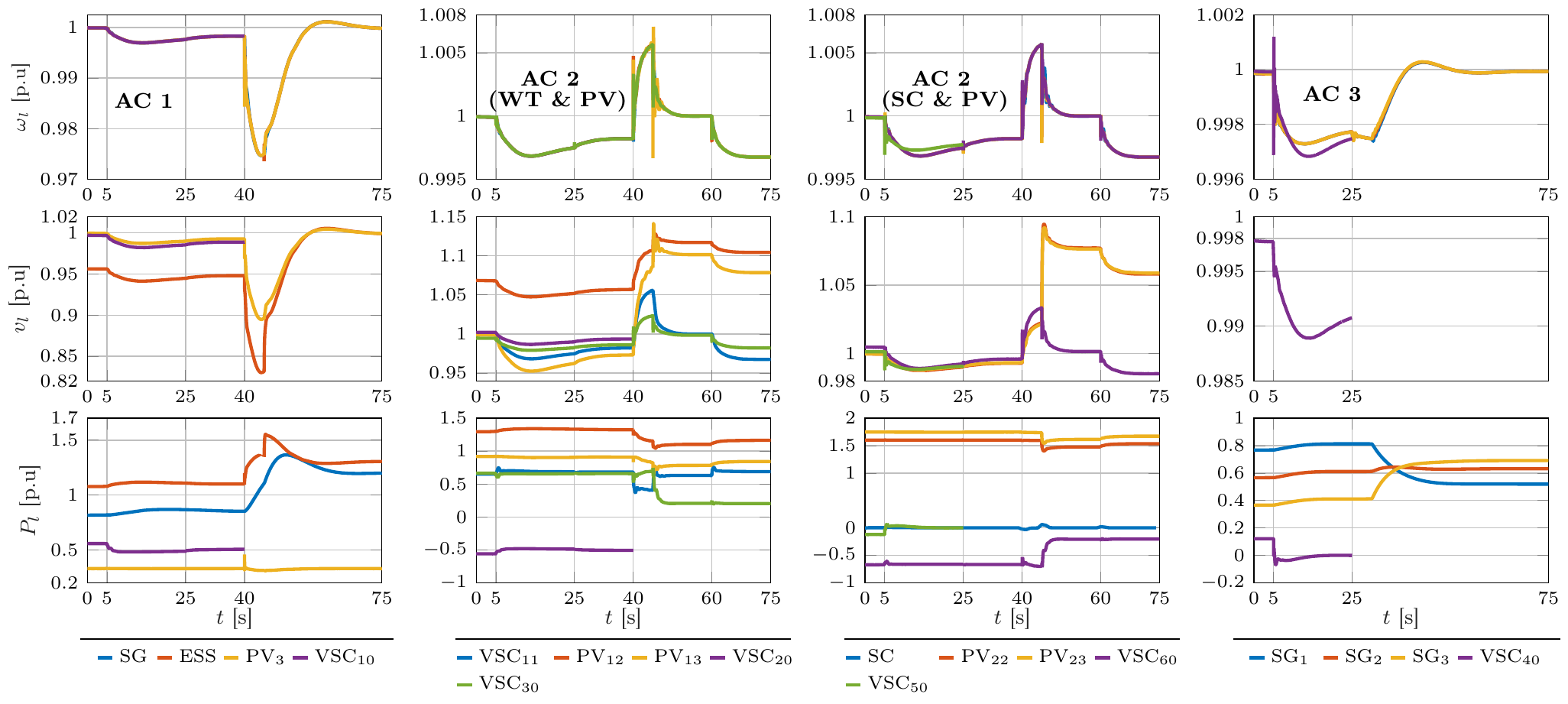} 
		\caption{Frequency $\omega_l$ (first  row), VSC dc voltage $v_l$ (second row), and injected power $P_l$ (third row) for AC~1 (first column), PMSG WT, and PV in AC~2 (second column), SC, and PV in AC~2 (third column), and AC~3 (forth column) during the sequence of events as in Tab.~\ref{table:event.seq}. Frequency and injected power are presented in p.u. with respect to the system base, while dc voltages are in p.u. with respect to the device base (e.g., $v_l^\mpp$ for a PV).		\label{fig:ac1-ac4_sim_results}}
	\end{figure*}

\begin{figure}[t!]
		\centering
		\includegraphics[width=\columnwidth]{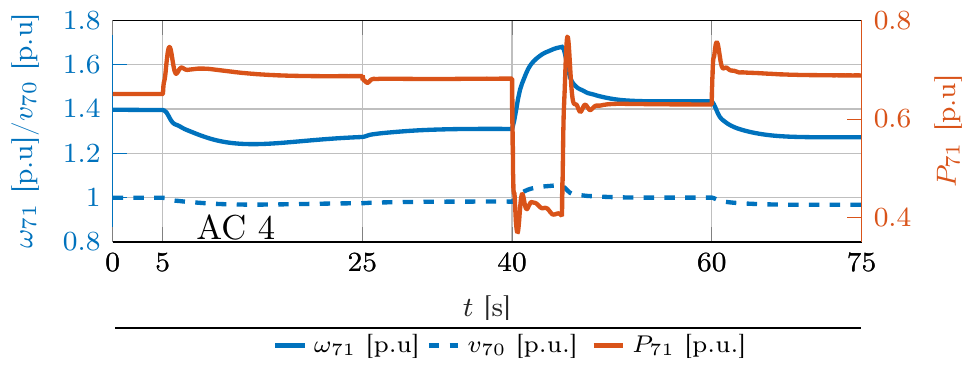} 
		\caption{PMSG rotor speed $\omega_{71}$, dc voltage $v_{70}$ of the WT VSCs, and generated power $P_{71}$.\label{fig:wt_sim_results}}
	\end{figure}

Figure~\ref{fig:ac1-ac4_sim_results} shows frequency, active power and dc voltage in the areas AC~1-AC~3, while Fig.~\ref{fig:wt_sim_results} shows PMSG WT frequency, dc voltage of the WT VSCs, and mechanical power of the WT (i.e., AC~4). Moreover, we emphasize that the $0.25$~p.u. load step is very large and pushes the system to the boundary of the normal operating range. Nonetheless, the system dynamics stay well-controlled. Moreover, as expected, the power imbalance propagates to all ac and dc subnetworks, and the curtailed  power sources provide primary frequency control according to their effective droop coefficients. After the load step at $t=5$~s, the PV systems operate at the MPP and do not change their power output, even though the dc voltage slightly varies around the MPP voltage. This verifies that the sensitivity of the PV power generation to dc voltage variations is approximately zero around the MPP (see  Fig.~\ref{fig:operating_points}). Moreover, as the load is increased, dual-port GFM control immediately adjusts the voltage phase angle of renewables approximately tracking the MPP to ensure that the dc voltage is well controlled. Therefore, only a small and brief transient is observed in their power injections. This  illustrates that  dual-port GFM control \eqref{eq:control.law} can exhibit conventional ac-GFL/dc-GFM functions without switching between GFM and GFL control modes. We stress that operating at $v_l<v_l^\mpp$ can be prevented by including a dc voltage limiter that becomes active when $v_l<v_l^\mpp$~\cite[Fig.~10]{CLJ19}.

\begin{table}[htb!]
	\centering			
	\caption{Sequence of events \label{table:event.seq}}	
	\resizebox{1\columnwidth}{!}{
		\setlength\tabcolsep{4pt}
		\bgroup
		\def\arraystretch{0.9}%
		\begin{tabular}{c c c c c c c c}
			\toprule
			$t$ & 0~s &	5~s & 25~s & 30~s & 40~s & 45~s & 60~s \\ \cmidrule{2-8}
			\multirow{2}{*}{\rotatebox{90}{event}} &nominal	&0.25~p.u. & DC~3 & AC~3 & DC~1 &  AC~2 & 0.125~p.u. \\ 
			& &	b37 & disconnect & redispatch & disconnect & redispatch & b16 $\&$ b29 \\ 	\bottomrule
	\end{tabular}
\egroup	}
\end{table}
	\begin{table}[htb!]
	\centering			
	\caption{Redispatched setpoints [p.u.]} \label{table:redisp.st}
	\resizebox{1\columnwidth}{!}{
		\setlength\tabcolsep{5pt}
		\bgroup
		\def\arraystretch{0.75}
		\begin{tabular}{c c c c c}
			\toprule
			\multirow{2}{*}{\rotatebox{0}{AC~1}} & \multicolumn{4}{c}{$P^\text{r}_{1}=1.2$,}\\
			&	$P^\text{r}_{2}\approx1.35$, & $k^\text{r}_{\g,2}\approx 1.16$,  & $k^\text{r}_{\omega,2}\approx 0.23$,& {$\kappa_{P,l}=5\%$,} \\  \midrule \midrule
			\multirow{5}{*}{\rotatebox{0}{AC~2}} &
			$P^\text{r}_{12}\approx86.9 \% P_{12}^\mpp$, & $k^\text{r}_{\g,12}\approx 4.2$,  & $k^\text{r}_{\omega,12}\approx 0.21$,& \multirow{4}{*}{$\kappa_{P,l}=5\%$,} \\  
			& $P^\text{r}_{13}\approx98.1 \% P_{13}^\mpp$,& $k^\text{r}_{\g,13}=2.1$, & $k^\text{r}_{\omega,13}\approx 0.11$, &  \\
			&  $P^\text{r}_{22}\approx 89.2\% P_{22}^\mpp$,  & $k^\text{r}_{\g,22}\approx 2.7$, & $k^\text{r}_{\omega,22}\approx 0.13$, &  \\
			&  $P^\text{r}_{23} \approx 87.7\% P_{23}^\mpp$,  &$k^\text{r}_{\g,23} \approx 2.8$, & $k^\text{r}_{\omega,23}\approx 0.14$,  & \\
			&  $P^\text{r}_{70}\approx89.1 \% P_{70}^\mpp$,& $k^\text{r}_{\g,70}=0.67$,&/ & {$\kappa_{P,70}\approx 3\%$,}\\\midrule \midrule
			\rotatebox{0}{AC~3} &
			\multicolumn{4}{c}{\begin{tabular}{c c c} $P^\text{r}_{31}=0.52$, & $P^\text{r}_{32}=0.63$, &	$P^\text{r}_{33}=0.69$
			\end{tabular}}\\ \bottomrule 
	\end{tabular}
\egroup	}
\end{table} 

Moreover, we use a loss of two HVDC lines as an extreme and unlikely contingency to illustrate that the system stays well-behaved even during severe events. After the loss of DC~3, the frequency in AC~1 and AC~2 rises, and the frequency in AC~3 drops since AC~3 was importing power. However, there are enough power reserves in AC~3 to support its load. Hence, after redispatching the power generation in AC~3, the frequency settles to its nominal value. On the other hand, after the loss of DC~1, the frequency in AC~1 significantly drops since power was imported via the HVDC link. Hence, in real-world scenarios, under frequency load shedding would be initiated in AC~1. However, for simulation purposes, we allow for very large frequency deviations and redispatch the system such that the SGs and ESS supply the load. Moreover, in contrast to the classical PLL-based ac-GFL/dc-GFM MPPT control, despite the large frequency deviation, the proposed control strategy reliably keeps the PV system in area AC~1 at the MPP. Finally, after both HVDC links are disconnected, the response of the converter-dominated system AC~2 is still well-behaved. After the loss of the DC~1, due to the relatively large mismatch of power production and consumption in AC~2, the frequency in AC~2 rises significantly. The proposed control responds to the power imbalance by reducing the renewable generation, i.e., exhibits conventional ac-GFM/dc-GFL grid support functions. Hence, to operate the system at the nominal frequency we redispatch ($t=45$~s) the renewable generation in AC~2 to a heavily curtailed operating point, resulting in nominal operation with significant reserves. The load steps occurring at $t=60$~s further test the system AC~2 containing only renewable generation, HVDC, and an SC. Even in this case, the system dynamics stay well-behaved. Furthermore, power imbalances propagate throughout the system, and the power sources share the additional load according to their effective droop coefficients $\kappa_P$.
	\subsection{Frequency RoCoF and Nadir}
\begin{figure}[t!]
	\centering
	\includegraphics[width=1\columnwidth]{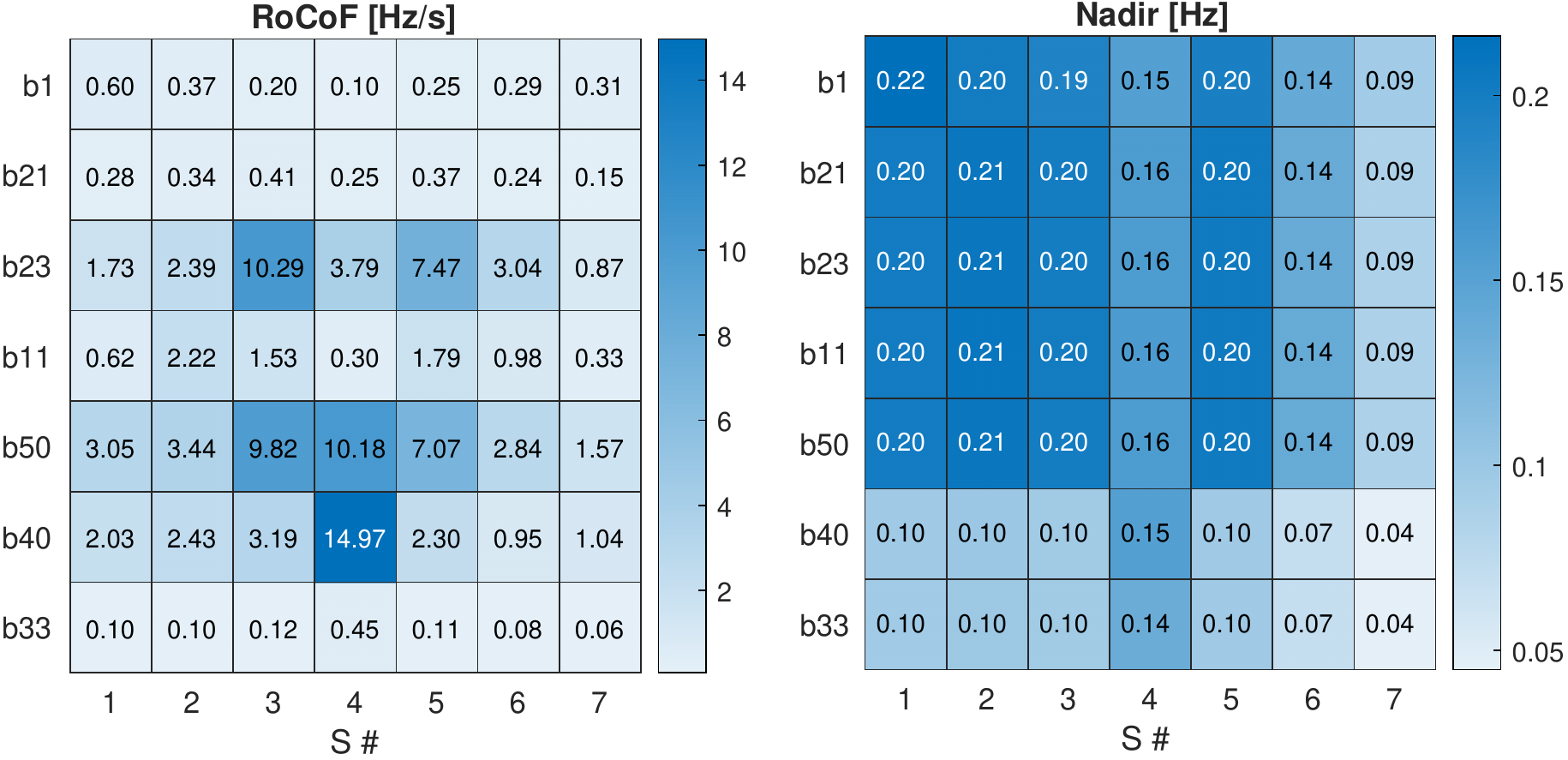} 
	\caption{Frequency RoCoF and Nadir for the load steps in Tab.~\ref{table:event.seq}. \label{fig:RoCoF_Nadir}}
\end{figure}

To compactly illustrate the system response to various load steps, we simulate seven different scenarios for the system in Fig~\ref{fig:grid}. The locations and values of the load steps occurring at $t=25$~s are given in Tab.~\ref{table:event.load-steps}. Figure~\ref{fig:RoCoF_Nadir} shows the RoCoF (calculated as the largest frequency change in the time window of $300$~ms) and the frequency nadir. We stress that the load steps are large and drive the system on the boundary of the nominal operating range. The largest RoCoF occurs at bus~40 during event \#4 (i.e., disturbance at bus~37) and the HVDC link (DC~3) is importing power from other areas. Notably, bus~40 corresponds to the HVDC VSC for which large RoCoF is not a concern. In contrast, during the same event, the RoCoF of the SG at bus~33 (i.e., close to the disturbance) stays within typical RoCoF limits of SGs. Finally, as discussed in Sec.~\ref{sec:ac.subnetwork.topology.cond.expanation}, Cond.~\ref{assump:identical.k.theta.gains} can be relaxed for point-to-point HVDC. To illustrate this aspect, the VSCs in DC~3 use significantly different gains $k_{\omega,l}$ and, as a result, the frequency nadir of devices in AC~3 and AC~1 and AC~2 differ significantly. To examine the frequency regulation performance further, an in-depth study of (i) the interplay between curtailment and the inertia response of renewables, as well as (ii) the ability of HVDC to transfer an inertia response between ac systems is seen as an interesting area for future work.
\begin{table}[htb!]
	\centering			
	\caption{Multiple load steps and their locations \label{table:event.load-steps}}
	\resizebox{0.7\columnwidth}{!}{
		\setlength\tabcolsep{4pt}
		\bgroup
		\def\arraystretch{0.73}%
		\begin{tabular}{c c c c c c c c}
			\toprule
			S \#	& 1  & 2 & 3 & 4 & 5 & 6 & 7 \\  \cmidrule{2-8}
			b7 & 0.25 & / & / & / & / & 0.0625  & 0.125 \\
			b15 & / &0.25 &  / & /& 0.083 & 0.0625 &  0.1 \\
			b29 & /& /& 0.25  & / & 0.167 & 0.0625 & / \\
			b39 & / & /& /& 0.25 & / & 0.1 &/ \\
			\bottomrule
	\end{tabular}
\egroup	}
\end{table}
\section{Conclusion and outlook} \label{sec:conclusion}
In this paper, we investigated end-to-end stability of universal dual-port GFM control hybrid dc/ac systems and renewable integration. We provided stability conditions and proved small-signal stability results for generic hybrid ac/dc systems containing different types of renewable generation, ac networks, and dc networks. Next, we interpreted the analytical stability conditions in the context of typical application scenarios and illustrated how to tune control gains to ensure stability of the overall system and meet steady-state response specifications. Adjusting control gains and renewable curtailment to meet desired transient performance specifications is seen as interesting topic for future work. Moreover, we illustrated how the proposed modeling and control framework can be used to model complex hybrid ac/dc systems. We used a detailed case study to illustrate (i) the ability of dual-port GFM control to provide grid support and approximately track the MPP of common renewables, and (ii) to show that the switching ripple affecting the dc voltage does not deteriorate stability. Additionally, we simulated a complex, large-scale system that includes standard legacy technologies and emerging technologies to illustrate our theoretical results and claims. While these results are encouraging, further work is needed to understand aspects such as current limiting and fault ride through that are well understood for GFL control but require further study for dual-port GFM control. A detailed study of single-stage and two-stage PV systems under time-varying irradition is seen as interesting area for future work.

\appendix

\subsection{Interconnection matrices}\label{app:intercon}
The interconnection of SMs and power sources with $k_{\g,l}>0$ are modeled by $\mc I_{\fr,\ac} \in \{0,1\}^{|\mc N_\fr| \times |\mc N_\ac|}$, i.e., $\{\mc I_{\fr,\ac}\}_{(i,j)}=1$ if the power source  $i \in \mc N_\fr$ is connected to the SM $j \in \mc N_\ac$ and $\{\mc I_{\fr,\ac}\}_{(i,j)}=0$ otherwise. Similarly, $\mc I_{\fr,\dc} \in \R^{|\mc N_\fr| \times| \mc N_\dc| }$ models the connection of dc nodes to dc power source with $k_{\g,l}>0$, i.e., $\{\mc I_{\fr,\dc}\}_{(i,j)}=1$ if $j \in \mc N_\fr$ is connected to $i\in \mc N_{\dc}$ and $\{\mc I_{\fr,\dc}\}_{(i,j)}=0$ otherwise. Analogously, the matrices ${\mc I}_{\zs,\ac}\in \{0,1\}^{ | {\mc N}_\zs| \times |\mc N_\ac| }$ and ${\mc I}_{\zs,\dc}\in \{0,1\}^{ |{\mc N}_\zs|  \times |\mc N_c| }$ model the interconnection between SMs and power sources ${\mc  N}_\zs$ and dc nodes and power sources ${\mc  N}_\zs$. Finally, the matrix $\mc I_\pv \in \{0,1\}^{|\mc N_\pv| \times |\mc N_\dc|}$ (resp. $\mc I_\w \in \{0,1\}^{|\mc N_\w| \times |\mc N_\ac|}$) models connections between dc buses and PV (resp. SMs and WTs). We also define $\mc I_\ac \in \{0,1\}^{|\mc N_\ac| \times |\mc N_\ac \cup \mc N_\cc|}$ and $\mc I_\ad \in \{0,1\}^{|\mc N_\cc| \times |\mc N_\cc \cup \mc N_\ac|}$ to extract machine and converter angles from the phase angle vector $\theta$, e.g., $\mc I_\ac \theta$ is the vector of all machine angles and $\mc I_\ad \theta$ collects all converter phase angles. Similarly, $\mc I_\da \!\in\! \{0,1\}^{|\mc N_\cc| \times |\mc N_\cc \cup \mc N_\dc|}$, and $\mc I_\dc \!\in\! \{0,1\}^{|\mc N_\dc| \times |\mc N_\cc \cup \mc N_\dc|}$ extract the dc voltages of VSCs and dc nodes from the vector $v$. 

\subsection{Proofs}\label{app:proof}

\begin{lemma}\label{lemma:help}	For all $i \in \N_{[1,N_\dc]}$ and $l \in \mc N_\da^i$, Cond.~\ref{lemma:more.conservative.gains.assumption} implies that
	$ 
	\sum\nolimits_{(l,k) \in \mc E_{\dc}^i} g^{\dc,i}_{lk} e_l +\sum\nolimits_{(l,k) \in \mc E_\dc^i \cap (\mc N_\da^i \times \mc N_\da^i)} g^{\dc,i}_{lk}\sqrt{e_l e_k}  <4 k^i_\omega
	$
	holds with $e_l\coloneqq k_{p,l}/c_l \in \R_{>0}$.
\end{lemma}

\textit{Proof of Lemma~\ref{lemma:help}:} First, we note that  $r^{\dc}_{\text{eq},l} = 1/\sum_{(l,k) \in  \mc E_\dc^i}g^{\dc}_{lk}$. Then, it directly follows from Cond.~\ref{lemma:more.conservative.gains.assumption} that $2\sum_{(l,k)\in \mc E_\dc^i}g^{\dc,i}_{lk}/(4 k^i_\omega) < c_l/k_{p,l}=e^{-1}_l >0 $ for all $l \in \mc N_\da^i$, i.e, $\max_{l \in \mc N_\da^i} e_l  < 4 k^i_\omega / (2\sum_{(l,k)\in \mc E_\dc^i}g^{\dc,i}_{lk})$. The Lemma directly follows from $\max_{\xi \in \mc N_\da^i} e_\xi  (\sum_{(l,k) \in \mc E_{\dc}^i} \! \! g^{\dc}_{lk}  +\!\sum_{k \in \mc N_\da^i} \! \! g^{\dc}_{lk}) \leq 2 \max_{\xi \in \mc N_{\da}^i} e_\xi \sum_{(l,k)\in \mc E_\dc^i}g^{\dc,i}_{lk} < 4 k^i_\omega$.\hfill \IEEEQED

	\textit{Proof of Prop.~\ref{prop:derivative.expression}}
	It can be verified that $\ddt V = \tfrac{1}{2}x_\delta^\T( \mc M A +  A^\T \mc M )x_\delta$. By substituting $\mc M$ and $A$ and performing elementary algebraic manipulations, it follows that $\ddt V=- \tilde{x}_\delta^T \mc V \tilde{x}_\delta -(\mc I_\w \omega_\delta)^\T K_\w \mc I_\w \omega_\delta-\tfrac{1}{2} v_\delta^\T(\tilde{K}_\omega \Xi+ \Xi \tilde{K}_\omega) v_\delta$.  By definition,  $K_\w\!\succ\!0$, $K_\pv\!\succ\!0$ and $\tilde{K}_{\omega}\!\succ\!0$.
	Then, using the Schur complement $\mc V\succeq 0$ holds if and only if 
	$	B_\dc^i {\mc W_\dc^i}^{\frac{1}{2}} \big( k_\omega^i I_{|\mc E_\dc^i|} - \tfrac{1}{4} {E^i}^\T E^i \big) {\mc W_\dc^i}^{\frac{1}{2}} {B_\dc^i }^\T \succ 0
	$ for all $ i \in \N_{[1,N_\dc]}$ and $E^i=\diag\{\sqrt{e_l}\}_{l=1}^{|\mc N_\cc^i|} \mc I_\da^i B_\dc^i {\mc W_\dc^i}^{\frac{1}{2}}$. Applying Gershgorin's theorem we obtain 
	$\lambda_{\max} \big({E^i}^\T \! E^i \big) \!\leq\! \max_{l \in \mc N_\da^i} \sum\nolimits_{(l,k) \in \mc E_{\dc}^i} g^{\dc,i}_{lk} e_l +\sum\nolimits_{k \in \mc N_\da^i} g^{\dc,i}_{lk}\sqrt{e_l e_k}$.
	Finally, using Lemma~\ref{lemma:help} it can be verified that Cond.~\ref{lemma:more.conservative.gains.assumption} ensures $\tilde{\mc V}\succeq 0$ and the proposition follows.\hfill \IEEEQED  
	
	\textit{Proof sketch of Prop.~\ref{prop:max.inv.set}:}
	Comparing \eqref{eq:control.law.lossless} and \eqref{eq:control.law.pbdp} and  with $m_{p,l}=\frac{k_{p,l}}{C_l}$, the difference between universal dual-port GFM control \eqref{eq:control.law} applied to the VSC model \eqref{eq:converter.model} and the power-balancing dual-port GFM control in~\cite{SG21} is the term $-\frac{k_{p,l}}{C_l}P_{\delta,\dc,l}$. This results in the term $K_p \mc I_{\da}C^{-1} L_\da v_\delta$ in \eqref{eq:matrix.A}. Using the same steps as in the proof of \cite[Prop. 2]{SG21} this term vanishes when $x_\delta$ is restricted to the invariant set $\bar{\mc S}$ and the proof of Prop.~\ref{prop:max.inv.set} follows from the proof of \cite[Prop. 2]{SG21}. \hfill \IEEEQED
	\textit{Proof of Prop.~\ref{prop:ss.sync.freq}:} Letting $\ddt v_{\delta,l}=0$ and combining \eqref{eq:converter.model}, \eqref{eq:pure.dc.node}, and \eqref{eq:control.dc.sources}, results in $(\mc I_\ad P_\ac,\mathbbl{0}_{|\mc N_\dc|})\!= \!P_{d_\dc}\!+\!(L_\dc+K_{\g,\dc}^\prime) (\mc I_{\da} v_\delta, \mc I_{\dc} v_\delta)$ with diagonal matrix $K_{\g,\dc}^\prime\!\coloneqq\! \mc I_\dc^\T (\mc I_\pv^\T\! K_\pv \mc I_\pv \! +  \!\mc I_{\fr,\dc}^\T\! K_\g \mc I_{\fr,\dc}) \mc I_\dc$. Using $\gc \! \coloneqq \! \max_{(l,k) \in \mc E_\dc} g^\dc_{l,k}$, we introduce the normalized dc network Laplacian $L_\dc^\prime\coloneqq\gc^{-1} L_\dc$. Next, we use Kron reduction \cite{DB2013} to eliminate all dc nodes to obtain $\mc I_\ad  P_{\ac,\delta}  = -(\gc\bar{L}_\dc + \Delta_\dc) \mc I_\da v_\delta - \bar{D}_\ad P_{d_\dc}$. From~\cite[Thm. III.6-3)]{DB2013} the matrices $\gc\bar{L}_\dc$, $\Delta_\dc$ (as a function of $\gc^{-1}$), and $\bar{D}_\ad$ correspond to the loop-less Laplacian, self-loops, and mapping of $P_{d_\dc}$ to the VSCs. In steady-state, \eqref{eq:control.law} implies $\mc I_\ad \omega_\delta\!=\!K_{\omega} \mc I_\da v_\delta$, i.e.,
	\begin{align}\label{eq:vscssacpower}
		\mc I_\ad  P_{\ac,\delta}  = -(\gc\bar{L}_\dc + \Delta_\dc) K^{-1}_{\omega} \mc I_\ad \omega_\delta - \bar{D}_\ad P_{d_\dc} 
	\end{align}
	Using the same arguments as in the proof of Prop.~\ref{prop:max.inv.set} and \cite[Prop.~8]{SG21} with $\ddt P_d\!=\!\mathbbl{0}_{n_d}$, Cond.~\ref{cond:topology.both} ensures synchronization to a synchronous frequency $\Omega_i$ for each ac subgrid $i\!\in\!\{1,\ldots,N_\ac\}$. Substituting \eqref{eq:vscssacpower} and $L \!\coloneqq\! B_\ac \mc W_\ac B_\ac^\T$ into the ac and ac/dc node steady-state equations results in 
	\begin{align*} 
		\! \begin{bmatrix}
			\!(\mc I_\w^\T  K_\w \mc I_\w \! +  \!\mc I_{\fr,\ac}^\T \! K_\g \mc I_{\fr,\ac}) \mc I_\ac\!\!\\
			(\gc\bar{L}_\dc + \Delta_\dc) K^{-1}_{\omega} \mc I_\ad
		\end{bmatrix}\!\omega_\delta\!=\!- L \theta_\delta \!-\! P_{d_\ac} \!\! \! - \! \mc I_\ad^\T \! \bar{D}_\ad\! P_{d_\dc}.
	\end{align*}
	Summing over the rows for each ac subnetwork results in $(\Delta \!+ \!\gc \bar{L}_\dc ^\prime ) \Omega_\delta \!=\!-\! \bar{P}_d$ where the vector $\Omega_\delta$ collect all $\Omega_{\delta,i}$. Next, $\{\mc B_\omega\}_{i,j}\! \! \in \!\! \{0,1\}^{|\mc N_\cc|\times N_\ac}$ models how VSCs are connected to the ac subnetwork (i.e., $\{\mc B_\omega\}_{i,j}\!\!=\!\!1$ if $i \! \! \in \! \!\mc N_\ad^j$). The matrix $\bar{L}_\dc^\prime \coloneqq \mc B_\omega \bar{L}_\dc K_\omega^{-1} \mc B_\omega^\T$ corresponds to the (loop-less) Laplacian matrix of a graph where nodes correspond to ac subnetworks and edges correspond to their interconnection through dc networks. The matrix $\Delta \! \coloneqq \! \diag\{\sum_{l\in \mc N_{\ac^\g}^i}\kappa_{P,l}^{-1}\}_{i=1}^{N_\ac} \! + \! \mc B_\omega \Delta_\dc K_{\omega}^{-1} \mc B_\omega^\T$ corresponds to the self-loops, and $\bar{P}_d \! \coloneqq \! \mc B_\omega\mc I_\ad^\T \! \bar{D}_\ad\! P_{d_\dc} \! +\! \diag_{i=1}^{N_\ac}\{\sum\nolimits_{l\in \mc N_\ac^i\cup \mc N_\ac^i}P_{d_\ac,l}\} \mathbbl{1}_{N_\ac} $ is a vector of total disturbances in each ac subnetwork. 
	We use $(\cdot)_\fr$ and $(\cdot)_\zs$ to differentiate between variables corresponding to the nodes with and without self-loops. Without loss of generality, we write $\blkdiag\{ \tilde{\Delta},\mathbbl{0}_{|\Omega_{\zs,\delta}|}\}(\Omega_{\fr,\delta},\Omega_{\zs,\delta})\!=\! -\! \gc \bar{L}_\dc^\prime (\Omega_{\fr,\delta},\Omega_{\zs,\delta})\! -(\bar{P}_{d,\fr},\bar{P}_{d,\zs})$. By Cond.~\ref{assump:onestab} $\tilde{\Delta}\succ 0$ is non-empty and collects all non-zero self loops from $\Delta$. 
	By applying Kron reduction \footnote{If $|\tilde{\Delta}|=1$, then $\tilde{\mc B}=\tilde{\mc B}^\T\!\coloneqq \!1$, $\tilde{\mc W}\!\coloneqq\!\tilde{L}$. Otherwise, $\tilde{B}$, $\tilde{\mc W}$ correspond to the incidence and edge weight matrices of the strictly-loopless Laplacian $\tilde{L}$.}
	to remove variables $\Omega_{\zs,\delta}$,  we obtain $(\tilde{\Delta} +\gc \tilde{L}) \Omega_{\fr,\delta} =\!-(\bar{P}_{d,\fr} \!+\! \tilde{D}_d \bar{P}_{d,\zs})$. From~\cite[Thm. III.6-3)]{DB2013}, the matrices $\gc\tilde{L}$ and $\tilde{D}_d$ correspond to the loop-less Laplacian, and mapping of $P_{d,\zs}$ to the nodes with self-loops. Solving the previous equation for $\Omega_{\fr,\delta}$ and multiplying it with $\tilde{B}^\T$ (the incidence matrix of $\tilde{L}$) we have $\tilde{B}^\T\Omega_{\fr,\delta}\!=\!-\!\tilde{B}^\T(\tilde{\Delta} +\gc \tilde{L} )^{-1}(\bar{P}_d+\tilde{D}_{d}\bar{P}_{d,\zs})$. 
	Applying the Woodbury matrix identity to $(\tilde{\Delta} +\gc \tilde{L} )^{-1}$ and letting $\gc\! \! \rightarrow \!\! \infty$, we obtain $\lim_{\gc\rightarrow \infty} \tilde{\mc B}^\T \Omega_{\fr,\delta} \! = \!\mathbbl{0}_{|\Omega_{\fr,\delta}|}$, i.e., $\Omega_{\fr,\delta,\infty} \! \! \coloneqq  \! \! \lim_{\gc \rightarrow \infty} \Omega_{\fr,\delta} \! \! \in \! \! \Null\{\tilde{\mc B}^\T\} \! \! = \! \! \delta_\omega\mathbbl{1}_{|\Omega_{\fr,\delta}|}$, with $\delta_\omega \! \! \in \! \!  \R$. Using $\lim_{\gc \rightarrow \infty} \mathbbl{1}_{|\Omega_{\fr,\delta}|}^\T \tilde{\Delta}(\tilde{\Delta}\! +\! \gc\tilde{L})^{-1}\!=\!\mathbbl{1}_{|\Omega_{\fr,\delta}|}^\T $ we have $\delta_\omega\!\!=\!-(\mathbbl{1}_{n_d}^\T P_d) /(\sum\nolimits_{l \in \mc N_{\fr} \cup \mc N_{\pv} \cup \mc N)_\w } \kappa_{P,l}^{-1})$. Moreover, it can be verified that $\Omega_{\zs,\delta,\infty}\!=\!\mathbbl{1}_{|\omega_{\zs,\delta}|} \delta_\omega$. The proof directly follows from $\omega^\sst_{\delta,l}\! =\! \Omega_{\delta,i}$,  for all $l\! \in \! \mc N_\ac^i\cup \mc N_\ad^i$ with $ i \! \in \!  \{\!  1,\ldots N_\ac\! \}$.  \hfill\IEEEQED

	\bibliographystyle{IEEEtran}
	\bibliography{IEEEabrv,bib_file}
	
%
	
\end{document}